\documentclass[a4paper,11pt]{article}
\usepackage[utf8]{inputenc}
\usepackage[DIV=12]{typearea}\usepackage{ragged2e}
\usepackage{calligra,amsmath,amsfonts,bbm,mathrsfs,amssymb}
\usepackage{slashed,cancel}
\usepackage{cite}
\usepackage{bm} 
\usepackage{hyperref}
\usepackage[dvipsnames]{xcolor}
\usepackage{colortbl}
\definecolor{verde}{cmyk}{0.92,0,0.59,0.25}
\definecolor{rossos}{cmyk}{0,1,1,0.55}
\hypersetup{colorlinks=true,urlcolor=rossos,anchorcolor=Blue,citecolor=verde,filecolor=verde,linkcolor=Blue,menucolor=Blue,linktocpage=true,pdfproducer=medialab}
\usepackage{graphicx}
\usepackage{cleveref}
\usepackage{bm}
\usepackage{mathdots}
\usepackage{color}
\usepackage{dsfont}
\usepackage{slashed}
\usepackage{physics}
\usepackage{xcolor}
\usepackage{subfigure}
\usepackage[normalem]{ulem}
\usepackage{comment}
\usepackage{dsfont}
\usepackage{lipsum}
\usepackage{mathtools}
\usepackage{url}
\usepackage{dcolumn}
\usepackage{booktabs}
\numberwithin{equation}{section}
\definecolor{blus}{cmyk}{1,1,0,0.6}

\newcommand{\email}[1]{\href{mailto:#1}{\tt #1}}

\definecolor{Nice}{HTML}{00B5BE}

\newcommand{\bea}{\begin{eqnarray}}
\newcommand{\eea}{\end{eqnarray}}
\newcommand{\beq}{\begin{equation}}
\newcommand{\eeq}{\end{equation}}

\newcommand{\X}{{\cal X}}

\newcommand{\U}{{\rm U}}

\renewcommand{\[}{\left[}

\begin{document} 

\renewcommand*{\thefootnote}{\fnsymbol{footnote}}
\begin{titlepage}

\vskip 1cm
\begin{center}
 {\LARGE \bf\color{Blue} Axion-Like Particles in Radiative \\ \vspace{0.3cm}  Quarkonia Decays}
\centering
\vspace{0.8cm}

{\bf Luca Di Luzio$^{a}$\footnote{\email{luca.diluzio@pd.infn.it}}, Alfredo Walter Mario Guerrera$^{a,b}$\footnote{\email{alfredowaltermario.guerrera@st.com}}
, \\ Xavier Ponce D\'iaz$^{a,b}$\footnote{\email{xavier.poncediaz@pd.infn.it}}, \bf Stefano Rigolin$^{a,b}$\footnote{\email{stefano.rigolin@pd.infn.it}}}\\[7mm]

$^a$~{\it Istituto Nazionale di Fisica Nucleare (INFN), Sezione di Padova, \\
Via F. Marzolo 8, 35131 Padova, Italy}\\[1mm]
$^b$~{\it Dipartimento di Fisica e Astronomia ``G.~Galilei'', Universit\`a di Padova,
 \\ Via F. Marzolo 8, 35131 Padova, Italy}\\[1mm]

\vspace{0.3cm}
\begin{abstract}
Radiative quarkonia decays offer an ideal setting for probing Axion-Like Particle (ALP) interactions. This paper provides a 
comprehensive review of ALP production mechanisms through the $e^+ e^- \to \gamma\,a$ process at B- and Charm-factories, 
alongside an analysis of potential ALP decay channels. We derive constraints on ALP couplings to Standard Model (SM) fields, 
based on recent experimental results on quarkonia decays by the Belle II and BESIII collaborations. 
The analysis distinguishes between ``invisible'' and ``visible'' ALP decay scenarios. The ``invisible'' scenario, characterised 
by a mono-$\gamma$ plus missing-energy signature, enables stringent limits on ALP-photon and ALP-quark ($b$ or $c$) couplings. 
Moreover, extensive research at flavour factories has explored various ``visible'' ALP decays into SM final states, which 
depend on a larger set of ALP-SM couplings. To streamline the ``visible'' ALP scenario, we introduce additional theoretical 
assumptions, such as universal ALP-fermion couplings, or we adopt specific benchmark ALP models, aiming to minimise the 
number of independent variables in our analysis.
\end{abstract}

\thispagestyle{empty}

\end{center}
\end{titlepage}

\setcounter{footnote}{0}

\pdfbookmark[1]{Table of Contents}{tableofcontents}
\tableofcontents

\renewcommand*{\thefootnote}{\arabic{footnote}}

\flushbottom
\setcounter{page}{1}

%%%%%%%%%%%%%%%%%%%%%%%%%%%%%%%%%%%%%%%%%%%%%%%%%%%%%%%%%%%%%%%
%%%%##################
\section{Introduction}
%%%%##################

Axion-Like Particles (ALPs) are a common feature of many extensions of the Standard Model (SM) of particle physics. Their 
lightness can be naturally justified if they are endowed with a pseudo-shift symmetry associated with the spontaneous breaking 
of an underlying global symmetry. A paradigmatic example is given by the QCD axion \cite{Peccei:1977hh,Peccei1977,
Weinberg:1977ma,Wilczek:1977pj}, which arises as a pseudo Nambu-Goldstone boson of a global $\U(1)_{\rm PQ}$ symmetry, 
anomalous under QCD and spontaneously broken at the scale $f_a\gg v$, where $v$ denotes the electroweak scale. The main 
difference between the QCD axion and an ALP lies in abandoning the requirement that the only explicit breaking of the 
$\U(1)_{\rm PQ}$ symmetry arises from non--perturbative QCD effects, leading to the well known relation $m_a f_a\approx 
m_\pi f_\pi$. Therefore, allowing the ALP mass, $m_a$, and symmetry breaking scale, $f_a$, to be independent parameters gives 
rise to a more general setup which is often described in terms of an effective Lagrangian containing operators up to $d = 5$ 
\cite{Georgi:1986df}. The opportunity to look for ALPs with masses well above the MeV scale, whose couplings are not tightly 
constrained by astrophysical limits, opens up the possibility of probing ALP interactions 
at colliders \cite{Mimasu:2014nea,Jaeckel:2015jla,Brivio:2017ije,Bauer:2017ris,Mariotti:2017vtv,Bauer:2018uxu,Alonso-Alvarez:2018irt,Aloni:2019ruo,Gavela:2019cmq,Bruggisser:2023npd} 
and via 
a broad class of
rare processes, including 
flavour-violating \cite{Izaguirre:2016dfi,Gavela:2019wzg,Cornella:2019uxs,MartinCamalich:2020dfe,
Calibbi:2020jvd,Bauer:2021wjo,Guerrera:2021yss,Gallo:2021ame,Bauer:2021mvw,Jho:2022snj,Guerrera:2022ykl,DiLuzio:2023ndz} and 
CP-violating \cite{Marciano:2016yhf,DiLuzio:2020oah,DiLuzio:2023cuk,DiLuzio:2023lmd} observables. 

Another relevant class of processes to probe ALP interactions is provided by radiative quarkonia decays of the type $V \to \gamma a$ 
\cite{Wilczek:1977zn}, with $V = \Upsilon, J/\psi$ a vector boson composed of heavy quarks. Historically, those observables
played a fundamental role in ruling out the original Weinberg-Wilczek axion, see for instance Ref.~\cite{Davier:1986ps}. 
ALP searches at B- and Charm-factories span mass regions from a few MeV up to roughly $10\, $GeV. This ALP mass range 
is poorly constrained by astrophysical and cosmological probes, as well as by beam-dump experiments, making ALP searches 
via quarkonia decays essential in constraining the ALP parameter space. Moreover, recent new experimental results on 
quarkonia decays by the BESIII \cite{BESIII:2022rzz} and Belle II \cite{Belle-II:2020jti} collaborations motivate an 
updated and thorough phenomenological analysis, to constrain ALP couplings to the $b$ and $c$ quarks, alongside 
other ALP couplings, including those to photons. As argued e.g.~in Ref.~\cite{Merlo:2019anv}, the simultaneous presence 
of ALP couplings to quarks and photons, expected in general ALP setups, gives rise to new interesting phenomenological features. 

In this work, we revisit the production of ALPs at B- and Charm-factories via the process $e^+ e^- \to \gamma a$. Firstly, 
we update the $\Upsilon\to\gamma a$ analysis of Ref.~\cite{Merlo:2019anv}, in which the ALP predominantly decays into an 
invisible channel (hereafter denoted as ``invisible'' ALP scenario), thus providing a mono-$\gamma$ plus missing-energy 
signature,  and extend the previous analysis to the charmonium sector. Moreover, we consider the complementary case in 
which the ALP decays visibly into the detector via SM final states. This scenario, denoted as ``visible'' ALP, opens more 
experimental channels to look for, albeit at the cost of enlarging the ALP parameter space. To reduce the number of 
independent parameters in the ``visible'' ALP scenario we make further theoretical assumptions, such as universal ALP-fermion 
couplings or consider a couple of benchmark ALP models, inspired by standard QCD axion models.  

The paper is structured as follows. In Sec.~\ref{sec:alp_lagrangian} we discuss the general form of the ALP effective 
Lagrangian below the electroweak scale and introduce two benchmark ALP models that will be used in the following 
phenomenological analysis. In Sec.~\ref{sec:ALPprod} we provide the general framework for ALP production in the context of 
B- and Charm-factories, while in Sec.~\ref{sec:ALPdecay} we collect general formulae for ALP decays into SM final states. 
Sec.~\ref{sec:pheno} is devoted instead to the phenomenological analysis, setting constraints both on the ``invisible'' and 
``visible'' ALP scenarios. Our results are summarised in Sec.~\ref{sec:concl}, while Appendix \ref{app:diagonalization} 
focuses on a technical issue, that is the exact diagonalisation of the ALP-pion mixing, which is relevant for our analysis 
and could also hold broader interest. 

%%%%%%%%%%%%%%%%%%%%%%%%%%%%%%%%%%%%%%%%%%%%%%%%%%%%%%%%%%%%%%%%%%%%%%%%%%%%%%%%%%%%%%%%%%%%%%%%%%%%%%%%%%%%%%%%%%%%%%%%%%
\section{The ALP Lagrangian}
\label{sec:alp_lagrangian}
%%%%%%%%%%%%%%%%%%%%%%%%%%%%%%%%%%%%%%%%%%%%%%%%%%%%%%%%%%%%%%%%%%%%%%%%%%%%%%%%%%%%%%%%%%%%%%%%%%%%%%%%%%%%%%%%%%%%%%%%%%

The most general CP-conserving and flavour-diagonal effective Lagrangian, describing the ALP-SM particle interactions at 
energies below the electroweak scale is given by the following $d=5$ operators:
\beq
\delta\mathcal{L}_{a}^{\rm SM}
=  - \frac{\partial_\mu a}{2f_a} \sum_{f\neq t} c_{aff}  \,\bar{f}\gamma^\mu\gamma_5 f - 
\, c_{a\gamma\gamma}\,\frac{\alpha_{em}}{4\pi} \frac{a}{f_a} \, F_{\mu\nu}\widetilde{F}^{\mu\nu} - 
c_{agg}\, \frac{\alpha_{s}}{4\pi} \frac{a}{f_a} \, G_{\mu\nu}^a\widetilde{G}^{\mu\nu}_a \, , 
 \label{eq:lagrangian2}
\eeq
with $f$ running over all the SM fermions but the top-quark and $\widetilde{V}^{\mu\nu}\equiv\epsilon^{\mu\nu\alpha\beta} 
V_{\alpha\beta}/2$ (with $\epsilon^{0123} = 1$) the dual field strength. Note that when considering ALP scenarios one typically 
works in the limit in which the ALP decay constant, $f_a$, is much larger than the ALP mass, i.e.~$f_a \gg m_a$. All 
couplings in Eq.~\eqref{eq:lagrangian2} are defined at the scale of the experiment, i.e.~$c_{aXX}\equiv c_{aXX} \,
(\mu=m_{\Upsilon}, m_{J/\psi})$. We here assume that the heavy SM fields, i.e.~top, Higgs and weak gauge bosons, have been 
integrated out and their effects are already encoded in the low-energy couplings by matching the high- and low-energy 
effective Lagrangians at the electroweak scale, and then running the Wilson coefficients down to the scale of the experiment
(for details, see Ref.~\cite{Bauer:2020jbp}). 

Another commonly used notation, especially in experimental papers, incorporates the scale $f_a$ in the coupling's definition 
and introduces the following mass-dimensional couplings:
\beq
g_{aff} = \frac{c_{aff}}{f_a} \, , \qquad 
g_{a\gamma\gamma} = c_{a\gamma\gamma} \frac{\alpha_{em}}{\pi f_a} \, , \qquad 
g_{agg} = c_{agg} \frac{\alpha_{s}}{\pi f_a} \, .
\label{eq:oldcouplings}
\eeq
In particular, notice the different $\alpha_{em}\,(\alpha_{s})$ normalization of the ALP-photon (gluon) coupling in the 
two versions of the Lagrangian, with the loop-origin of the photon (gluon) coupling being highlighted in the notation of Eq.~\eqref{eq:lagrangian2}.

ALPs may also act as portals to a light-dark sector. In this case, additional ALP couplings are introduced. Assuming, for 
simplicity, the existence of a single light and neutral dark fermion $\chi$, the following term is added to the $d=5$ ALP 
effective Lagrangian:
\beq
\delta\mathcal{L}_{a}^{\rm DS} = -\frac{\partial_\mu a}{2f_a}\,c_{a\chi\chi}\,\bar{\chi}\gamma^\mu\gamma^5\chi \, , 
\label{eq:lagrangian3}
\eeq
with $c_{a\chi\chi}$ a coupling that can induce a sizeable ALP decay into invisible final states, whenever $m_a\gtrsim 2\,m_\chi$. 

%%%%%%%%%%%%%%%%%%%%%%%%%%%%%%%%%%%%%%%%%%%%%%%%%%%%%%%%%%%%%%%%%%%%%%%%%%%%%%%%%%%%%%%%%%%%%%%%%%%%%%%%%%%%%%%%%%%%%%%%%%
\subsection{Benchmark ALP models}
\label{sec:UVmodels}
%%%%%%%%%%%%%%%%%%%%%%%%%%%%%%%%%%%%%%%%%%%%%%%%%%%%%%%%%%%%%%%%%%%%%%%%%%%%%%%%%%%%%%%%%%%%%%%%%%%%%%%%%%%%%%%%%%%%%%%%%%

The effective field theory approach of Eq.~(\ref{eq:lagrangian2}) features several Wilson coefficients, which do not allow in 
general for a simple representation of the experimental constraints from quarkonia decays. Therefore, it is also useful to 
consider ultraviolet (UV) complete scenarios which provide non-trivial correlations among the Wilson coefficient, thus 
considerably reducing the ALP parameter space.  In the following, we will provide a couple of benchmark ALP frameworks, inspired 
by the canonical DFSZ \cite{Zhitnitsky:1980tq,Dine:1981rt} and KSVZ \cite{Kim:1979if,Shifman:1979if} axion models. Differently 
from the case of the QCD axion, we will assume here that the ALP mass is a free parameter. 
See also Ref.~\cite{Arias-Aragon:2022iwl} for related examples.  

\subsubsection*{DFSZ-QED ALP}

A useful benchmark scenario is provided by a variant of the DFSZ model with no QCD-PQ anomalous couplings, that we 
denote as DFSZ-QED. This is obtained by means of the following Yukawa Lagrangian, with the two Higgs doublets denoted as $H_q$ and $H_\ell$
\begin{align}
    -\mathcal{L}^{\textrm{DFSZ-QED}}_Y=
    Y_u \,\bar q_L u_R \, \tilde{H}_q+Y_d \,\bar q_L d_R \, H_q+Y_e \,\bar \ell_L e_R \, H_{\ell} +\rm{h.c.} \, .
\end{align}
The scalar potential is assumed to contain an operator of the type $H_q^\dagger  H_\ell\phi^{\dagger 2}$, with $\phi$ a SM-singlet complex scalar. This operator constrains the PQ charges of the scalar fields, generically denoted by $\X_{q,\ell,\phi}$ and normalised so that $\X_\phi=1$, to satisfy $\X_\ell-\X_q=2\X_\phi= 2$. Imposing the orthogonality of the PQ and hypercharge currents and defining $\tan\beta \equiv t_\beta=v_\ell/v_q$, with $v_\textrm{ew}^2 \equiv v_q^2 + v_\ell^2 = (246 \ \text{GeV})^2$, one obtains $\X_q = -2s_\beta^2$ and $\X_\ell=2c_\beta^2$. It can be readily verified that this model implies no QCD-PQ anomaly since SM quarks interact with the same Higgs doublet. The ALP couplings to SM fields can then be obtained following standard techniques (see e.g.~\cite{DiLuzio:2020wdo}) and, 
matching onto the notation of Eq.~(\ref{eq:lagrangian2}), one finds 
\begin{align}
\text{DFSZ-QED:} \qquad 
    c_{auu} &=-c_{add}=-2 s_\beta^2 \, ,  \quad  
    c_{aee} = - 2c_\beta^2 \, , \quad 
c_{a\gamma\gamma} = 6 \, , \quad
c_{agg} = 0 \, , 
        \label{eq:couplDFSZQED}
\end{align}
where the ALP-SM fermions couplings are understood to be 
universal.

\subsubsection*{KSVZ ALP}

As a second benchmark setup, we consider a KSVZ-like ALP model in which the PQ anomaly is carried by a new coloured heavy fermion, 
$\mathcal{Q}$, charged under the PQ symmetry, with Yukawa interaction $\overline{\mathcal{Q}}_L \mathcal{Q}_R \phi$, and 
$\phi$ a SM singlet field carrying unit PQ charge. After integrating out the new heavy fermion, assuming e.g.~that it 
transforms in the fundamental of colour, one obtains (see e.g.~\cite{DiLuzio:2020wdo})
\begin{align}
\text{KSVZ:} \qquad
    c_{auu} = c_{add} = c_{aee} = 0 \, , \quad 
c_{a\gamma\gamma} = 0 \, , \quad
c_{agg} = 1/2 \, , 
        \label{eq:couplKSVZ}
\end{align}
for the ALP-SM tree-level couplings.

%%%%%%%%%%%%%%%%%%%%%%%%%%%%%%%%%%%%%%%%%%%%%%%%%%%%%%%%%%%%%%%

\section{ALP production mechanisms}
\label{sec:ALPprod}

The ALP production mechanisms at B- and Charm-factories via the process $e^+ e^- \to \gamma a$ have been carefully 
analysed in Ref.~\cite{Merlo:2019anv}, to which we refer for most of the details. In the following, we will shortly recall 
the main features that will be relevant for the phenomenological discussion of the subsequent sections. 

The simplest way of producing an ALP in $e^+e^-$ colliders is via the non-resonant tree-level process $e^+e^-\to 
\gamma a$, that is dominated by the $s$-channel photon mediated diagram. The total non-resonant cross section, in the 
centre-of-mass rest frame, is given by \cite{Marciano:2016yhf}: 
\beq
\sigma_\mathrm{NR}(s) = \frac{\alpha_\mathrm{em}^3}{24\pi^2} \frac{c^2_{a \gamma\gamma}}{f_a^2}\, 
                        \left(1-\frac{m_a^2}{s}\right)^3 \, .
\label{eq:non_resonant}
\eeq
As we are interested in a relatively large ALP mass range (from few MeV up to about 10 GeV) the $m_a$ dependence is fully accounted for 
in Eq.~\eqref{eq:non_resonant}, while terms proportional to $m_e^2/s$ can be safely neglected at the energies relevant for flavour 
factories. Note that the non-resonant ALP-photon production via a $t$-channel electron contribution is usually neglected 
as suppressed by the electron mass: the underlying assumption here is that $c_{aee}/c_{a\gamma\gamma} \lesssim 10^2$ (or 
equivalently $g_{aee}/g_{a\gamma\gamma} \lesssim 10^4$).\footnote{The contribution coming from the $s$-channel exchange 
of an off-shell $Z$ boson, that would appear in the low-energy ALP 
Lagrangian as a $d=6$ operator is also typically neglected 
since it is suppressed by $s/M_Z^2 \ll 1$ at low energies. In this case the underlying assumption $c_{a\gamma Z}/c_{a\gamma\gamma} 
\lesssim 10^{2}$ is also implied.}

While the non-resonant contribution to ALP production in Eq.~\eqref{eq:non_resonant} is unavoidable in any $e^+ e^-$ 
experiment, the situation at B- and Charm-factories is more intricate since these experiments operate around 
specific resonances (i.e.~$\Upsilon(nS)$ or $J/\psi(nS)$). It is, therefore, crucial to correctly account 
for the resonantly enhanced contributions. Generalising the discussion in~\cite{Merlo:2019anv}, the resonant cross 
section for an ALP production via vector quarkonia resonances, $V=\Upsilon,J/\psi$, in the Breit-Wigner approximation 
reads:
\beq
\sigma_{\rm R}(s)=\sigma_\mathrm{peak} \frac{m_V^2 \Gamma_V^2}{(s-m_V^2)^2 +m^2_V\Gamma_V^2}\mathcal{B}(V\to \gamma a) \, , 
\label{eq:resonant}
\eeq
where $m_V$ and $\Gamma_V$ are, respectively, the mass and the width of the specific resonance and $\sigma_\mathrm{peak}$ 
is the peak cross section 
\beq
\sigma_\mathrm{peak} =\frac{12\pi \mathcal{B}(V\to e^+e^-)}{m_V^2} \, ,  \label{eq:peak}
\eeq
defined in terms of the leptonic branching fraction, $\mathcal{B}(V\to e^+e^-)$, experimentally determined for each 
different intermediate state~\cite{Workman:2022ynf}. The effective ALP couplings defined in Eq.~(\ref{eq:lagrangian2}) 
enter in the $\mathcal{B}(V\to a\gamma)$ branching fraction as \cite{Merlo:2019anv}:
%%%
\beq
\mathcal{B}(V\to\gamma a)=\frac{\alpha_\mathrm{em} Q_Q^2}{24}\frac{m_V\,f_V^2}{\Gamma_V\,f_a^2}
    \left(1-\frac{m_a^2}{m_V^2}\right)
    \left[c_{a\gamma\gamma}\,\frac{\alpha_\mathrm{em}}{\pi}\left(1-\frac{m_a^2}{m_V^2}\right)-2\,c_{aQQ}\right]^2 \, , 
\label{eq:branching_ratio}
\eeq
%%%
with $Q_Q$ and $c_{aQQ}$ the electromagnetic charge and the ALP-fermion couplings of the quarkonia valence quarks, i.e. 
$Q=c,b$ for the cases under consideration\footnote{Notice that the calculation of the hadronic matrix elements needed 
to estimate the $g_{aQQ}$ contribution is done within the approximation in which the two partons share exactly half of 
the meson momentum. See Ref.~\cite{Guerrera:2022ykl} and references therein for a detailed discussion of this point.}, 
while $f_V$ are the quarkonium decay constants, which can be obtained from \cite{ParticleDataGroup:2022pth}. 

Naively, one would expect the resonant contribution to always dominate by orders of magnitude over the non-resonant one. 
However, this expectation is incorrect in some of the cases under consideration, since the resonance widths are typically much 
narrower than the energy beam resolution, $\sigma_W$. Therefore, when the resonance is not fully resolved by the 
experiment, a convolution of the theoretical resonant contribution with a Gaussian spread function has to be performed 
(see e.g.~\cite{Eidelman:2016aih}) and an ``experimental" resonant cross section can be defined as:
\beq
\left< \sigma_{\rm R}(s)\right>_\mathrm{exp}=\frac{1}{\sqrt{2\pi}}\int dq \,\frac{\sigma_R(q^2)}{\sigma_W}\exp
\left[-\frac{(q-\sqrt{s})^2}{2\sigma_W^2} \right] \, .
\label{eq:smeareing}
\eeq
In the case of very narrow resonances, the previous result can be simplified to:
\beq
\left< \sigma_R(s)\right>_\mathrm{exp}= \rho\, \sigma_\mathrm{peak} \, \mathcal{B}(V\to\gamma a) \, , \qquad 
\rho = \sqrt{\frac{\pi}{8}} \, \frac{\Gamma_V}{\sigma_W} \, , 
\label{eq:smeareingNarrow}
\eeq
with the parameter $\rho$ accounting for the suppression of the ``experimental'' resonant cross section due 
to the non-negligible beam-energy spread. 

Depending on the experimental setup, ALP searches at flavour factories can be classified into three different categories, 
based on the different ALP production mechanisms. 
\paragraph{i) Non-resonant searches:} This scenario arises when the ALP is produced off-resonance or the resonance is 
very spread, as for example is the case of Belle II searches at the $\Upsilon(4S)$ resonance \cite{Belle-II:2020jti}. 
These searches are dubbed {\em non-resonant} as in this case the non-resonant cross section in Eq.~\eqref{eq:non_resonant} 
has to be used and therefore only a sensitivity to $g_{a\gamma\gamma}$ can be obtained from the ALP production.
\paragraph{ii) Resonant searches:} 
Excited quarkonia states can decay into lighter quarkonia resonances via pion emission, for example, $\Upsilon(2S) \to 
\Upsilon(1S) \pi^+ \pi^-$. By exploiting the kinematics of the final states, 
the lighter meson resonance can be fully reconstructed and its 
mono-$\gamma$ decay mode $\Upsilon(1S)\to \gamma \,a$ analysed, as done for example in Ref. \cite{Belle:2019iji}. These 
searches are dubbed in the following as {\em resonant} since they allow to directly probe $\mathcal{B}(V\to\gamma a)$ in 
Eq.~\eqref{eq:branching_ratio} and give information on a specific combination of $g_{a\gamma\gamma}$ and $g_{aQQ}$ couplings. 
\paragraph{iii) Mixed searches:} Experimental searches sometimes are performed at an energy that is around a specific 
(narrow) resonance $\sqrt{s} \approx m_{V(nS)}$, but without identifying it kinematically. From the above discussion 
it is clear that in this case, due to the larger beam energy uncertainties, the production mechanism depends on both the 
resonant and non-resonant contributions, being sensitive to a different $g_{a\gamma\gamma}$ and $g_{aQQ}$ combination, 
compared to the resonant searches. In the case under examination, the following simple prescription provides a good 
approximation of the production cross section \cite{Merlo:2019anv}:
\beq
\left< \sigma_{\rm mix} (s)\right>_\mathrm{exp} \approx  \sigma_{\rm NR}(s) + \left<\sigma_{\rm R}(s)\right>_\mathrm{exp} \,.
\label{eq:mixed}
\eeq
In Tab.~\ref{tab:RvsNR} an estimate of the ``experimental" resonant cross sections compared to the non-resonant ones 
is presented, for the relevant decay channels. The experimental energy spread of the $e^+ e^-$ beams at flavour factories 
can be typically taken in the range $\sigma_W\approx 2-5$ MeV \cite{BESIII:2009fln,Workman:2022ynf,Koiso:2013tgf,Ohnishi:2013fma}. 
This value is considerably larger than all the considered resonance widths, typically in the $20-100$ keV range, with the only 
exception of the $\Upsilon(4S)$ resonance, for which $\Gamma_\Upsilon(4S) = 20.5$ MeV. 
\begin{table*}[!t]
\centering
  \renewcommand{\arraystretch}{1.4} 
  %\footnotesize
\begin{tabular}{c|cc|cc|c}
$V(nS)$  & $m_V~[\mathrm{GeV}]$ &  $\Gamma_V~[\mathrm{keV}]$  & $\sigma_{\mathrm{peak}}$~[nb] & $\rho$  
& $\langle\sigma_{\mathrm{R}}\rangle_{\mathrm{exp}}/\sigma_{\text{NR}}|_{s=m_V^2}$\\
\hline
$J/\Psi(1S)$    & $3.096$  & $92.6$  & $91.2(5)\times 10^{3} $  & $31 \times 10^{-3}$ &  $ 0.92(1)$ \\
$\Upsilon(1S)$  & $9.460$  & $54.02$  & $3.9(2)\times 10^{3}$             & $6.1\times 10^{-3}$ &  $0.53(5)$ \\
$\Upsilon(2S)$  & $10.023$ &  $31.98$  & $2.8(2)\times 10^{3}$             & $3.7\times 10^{-3}$ &  $0.21(3)$ \\
$\Upsilon(3S)$  & $10.355$ & $20.32$  & $3.0(3)\times 10^{3}$              & $2.3\times 10^{-3}$ &  $0.16(3)$ \\ 
$\Upsilon(4S)$  & $10.580$ & $20.5\times 10^3$  &  $2.1(1)$ & $0.83$    & $3.0(3)\times 10^{-5}$ \\
\end{tabular}
\caption{\em ``Experimental'' cross sections at BESIII and Belle II for $e^+e^-\to V \to \gamma a$, compared 
to the non-resonant ones, $e^+e^-\to \gamma^\ast \to \gamma a$. Vanishing ALP couplings to $c$- and $b$-quarks have 
been assumed here.}  
\label{tab:RvsNR}
\end{table*}
%%%%%%%%%%%%%%%%%%%%%%%%
%
As it can be seen, for the $\Upsilon(4S)$ case the Gaussian smearing is practically ineffective, i.e.~$\rho \approx 1$ being 
$\Gamma_\Upsilon(4S) \gg \sigma_W$. However, at the same time, the broadness of the resonance largely reduces the resonant 
contribution, five orders of magnitude below the non-resonant one, due to the smallness of the corresponding $\sigma_\mathrm{peak}$ 
value. In all the other cases, instead, the smearing procedure 
strongly suppresses the naive Breit-Wigner theoretical 
expectation by roughly a factor of $10^{-3}$. In these cases, the ``experimental" resonant cross section is smaller than 
the non-resonant one, even if it can still contribute with numerically significant effects between 20\% and 50\% of 
the non-resonant one, which should therefore be taken 
into account when interpreting experimental searches. For illustrative 
purposes, the numbers reported in Tab.~\ref{tab:RvsNR} have been calculated assuming vanishing ALP-fermion couplings 
and, as a consequence, the $c_{a\gamma\gamma}$ dependence simply cancels out in the cross section ratio. In the general case, 
with more ALP couplings, results can slightly change. 

A detailed discussion regarding all the subtleties entering in the interpretation of the $e^+ e^- \to \gamma \,a$ production 
at the $\Upsilon(nS)$ resonances has been performed in Ref.~\cite{Merlo:2019anv}, to which the interested reader is referred.

\section{ALP decay channels}
\label{sec:ALPdecay}

Being interested in ALP production from radiative quarkonia decays, in the following, we will focus on ALP masses 
$m_a \lesssim 2 \,m_q$, with $q=c,b$ depending on whether $V=J/\Psi$ or $\Upsilon$. The total ALP decay width in SM 
particles reads:
\beq
\Gamma(a\to \mathrm{SM}) = \frac{\alpha_{\textrm{em}}^2 }{64 \pi^3 f_a^2}\abs{c_{a\gamma\gamma}^\textrm{eff}}^2 m_a^3 + 
     \sum_{f} \frac{N_c^f \abs{c_{aff}}^2}{8\pi f_a^2}m_a m_f^2
     \sqrt{1-\frac{4 m_f^2}{m_a^2}}\,\Theta\left(m_a - 2m_f \right) +\Gamma_{a\to\textrm{lh}}\,,
\label{eq:gamma_tot}
\eeq
where $\Theta(x)$ is the Heaviside step function, $N_c^f=1(3)$ for leptons (quarks) and $f=\{e,\,\mu,\,\tau,\,c,\,b\}$ 
extends to all kinematically accessible fermions, except for light quark flavours which are included in the the light 
hadronic decay width, $\Gamma_{a\to\textrm{lh}}$, that is provided in Eq.~\eqref{eq:lighth}, while 
the effective ALP couplings to photons is defined below in Eq.~\eqref{eq:caeffgaga}.

In the $m_a\sim\mathcal{O}(\textrm{GeV})$ region, ALP decays into light hadrons become at the same time 
relevant and difficult to compute, requiring different approaches depending on the value of $m_a$. While 
for $m_a \lesssim 1\,$GeV one can use chiral perturbation theory to predict exclusive hadronic ALP decays, for $m_a\gtrsim 
2\,$GeV one 
can rely on the quark-hadron duality \cite{Poggio:1975af,Shifman:2000jv} to compute the inclusive 
ALP decay rate into hadrons. However, in the $m_a \in \left[1,2\right]\, $GeV region, one lays outside the applicability range 
of both chiral perturbation theory and perturbative QCD. For this reason,
the 1-2 GeV ALP mass range 
will be conservatively excluded when deriving bounds on the ALP-SM couplings\footnote{See however Ref.~\cite{Aloni:2018vki} 
for a comprehensive data-driven approach in order to deal with hadronic observables in 
this ALP mass range.}. The ALP decay width into light hadrons, in the two $m_a$ regions considered, is then given 
by~\cite{Spira:1995rr,Bauer:2017ris}:

\beq
\Gamma_{a\to\textrm{lh}}=
\begin{dcases} \,
\frac{m_a m_\pi^4}{6144 \pi^3 f_\pi^2 f_a^2}\abs{c_{a\pi}}^2\left( g_{00}\left(\frac{m_\pi^2}{m_a^2}\right) + 
g_{+-}\left(\frac{m_\pi^2}{m_a^2}\right)\right) \quad & m_a\lesssim 1\,\textrm{GeV}\, ,  \\
\, \frac{\alpha_s^2}{8\pi^3 f_a^2}\abs{ c_{agg}^{\textrm{eff}}}^2m_a^3\left(1+\frac{\alpha_s}{4\pi}\frac{291 -14\,n_q}{12}\right) \, 
\qquad  & m_a \gtrsim 2\,\textrm{GeV}\, ,
\end{dcases} 
\label{eq:lighth}
\eeq
where $n_q$ is the number of active quark flavours and we have introduced the ALP-pion and ALP-gluon effective couplings, defined respectively by 
\bea
c_{a\pi} & = & -\left(2c_{agg}\frac{m_d-m_u}{m_d+m_u}+c_{auu}-c_{add}\right)  \, , \label{eq:capi} \\ 
c_{agg}^{\textrm{eff}} & = & c_{agg}+\frac{1}{2}\sum_{q\neq t} c_{aqq} B_1\left(\frac{4m_q^2}{m_a^2}\right) \, , \label{eq:cagg}
\eea
with the loop function $B_1(\tau)$ given by 
\beq 
B_1(\tau)=1-\tau f(\tau)^2 \qquad \textrm{with}\qquad f(\tau)=\begin{dcases}
    \,\arcsin(\frac{1}{\sqrt{\tau}}) \quad &\tau \geq 1 \, , \\
    \,\frac{\pi}{2}+\frac{i}{2}\log{\left(\frac{1+\sqrt{1-\tau}}{1-\sqrt{1-\tau}}\right)} \quad &\tau <1 \, , 
\end{dcases}
\eeq
respectively below and above the quark production threshold. The terms proportional to the functions $g_{00}$ and $g_{+-}$ 
in Eq.~\eqref{eq:lighth} correspond respectively to the three-body decays of $a\to 3\pi^0$ and $a\to \pi^+\pi^- \pi^0$, 
while other three-body decays containing electrons or photons are suppressed by powers of $\alpha_{em}$. The explicit 
expression of the functions $g_{00}$ and $g_{+-}$ can be found in Ref.~\cite{Bauer:2017ris}, where the reader is referred 
for additional details. It should be mentioned 
that the chiral description of $a \to 3\pi$ processes breaks 
down just above kinematical threshold \cite{DiLuzio:2022tbb}, so although we expect these decay widths to be $\mathcal{O}(1)$ 
correct, a more refined description of the ALP width below the GeV scale would require techniques beyond 
standard chiral perturbation 
theory (see e.g.~\cite{DiLuzio:2022gsc} for a related example). 

Finally, in Eq.~(\ref{eq:gamma_tot}) the effective photon coupling, $c_{a\gamma\gamma}^{\textrm{eff}}$, has been defined 
to include all possible one-loop contributions. Again, this coupling needs to be defined differently in the two considered 
ALP mass regions. Neglecting the subleading top-quark and $W$ loops one obtains
\cite{Bauer:2017ris,Bauer:2020jbp}
\beq
c_{a\gamma\gamma}^{\textrm{eff}} = 
\begin{dcases} \vspace{-0.1cm}
c_{a\gamma\gamma}-1.92\, c_{agg}+\frac{f_a}{f_\pi}U_{a\pi}+\sum_{f=\ell,c,b} N_c^f Q_f^2 c_{aff} B_1(\tau_f) \quad  & 
m_a \lesssim 1\,\textrm{GeV} \, , \\
\vspace{-0.1cm}\, c_{a\gamma\gamma}+\sum_{f\neq t} N_c^f Q_f^2 c_{aff} B_1(\tau_f) \qquad\qquad & m_a\gtrsim 2\,\textrm{GeV} \,, 
\end{dcases}
\label{eq:caeffgaga}
\eeq
with $Q_f$ the fermions' electromagnetic charge. 
An approximate expression for the ALP-pion mixing term, $U_{a\pi}$, 
valid for $\abs{m_\pi^2-m_a^2}/\text{max}\{m_\pi^2,m_a^2\} \gg f_\pi/f_a$, reads \cite{Bauer:2017ris,Bauer:2020jbp,DiLuzio:2022tbb}
\beq
U_{a\pi} \approx 
\frac{m_a^2}{m_\pi^2-m_a^2}\frac{f_\pi}{2f_a}c_{a\pi} \, . 
\eeq
More details on the ALP-pion diagonalisation procedure and the exact expression for the mixing angle are reported for completeness 
in Appendix \ref{app:diagonalization}.

In Fig.~\ref{fig:branchingratios} we display the branching ratio $\mathcal{B}(a \to \text{SM})$, based on Eq.~\eqref{eq:gamma_tot} 
and assuming, for the sake of illustration, all the Wilson coefficients to be $c_{aXX}=1$ and $f_a=1\,$TeV. In this way, 
one can compare the relative strengths of the ALP decay to SM final states. In the sub-GeV ALP mass region, 
the ALP decays preferably in electrons and muons, except in the proximity of the maximal mixing region, i.e.~$m_a=m_\pi$, 
where the ALP-photon coupling gets largely enhanced, while for $m_a \gtrsim 2\, $GeV the larger ALP branching ratio is 
dominantly into light and heavy hadrons with the $\tau$ contribution is only slightly subdominant when kinematically allowed.
Note, however, that in the conventions of Eq.~\eqref{eq:lagrangian2}, the fine-structure constant $\alpha_{em}$ 
is factorised outside the ALP-photon coupling, $c_{a\gamma\gamma}$. Nonetheless, 
the non-decoupling nature of the anomalous terms may easily give rise to larger contributions, depending 
on the multiplicity of heavy electromagnetically-charged exotic fermions in the UV-completed model (see e.g.~\cite{DiLuzio:2016sbl,
DiLuzio:2017pfr}). If, instead, the alternative definitions of Eq.~\eqref{eq:oldcouplings} for 
the ALP-SM couplings are used, and $g_{aXX}=1$ is assumed, the ALP-photon branching ratio would be enhanced by a factor 
$\sim 10^{5}$, thus resulting in the dominant contribution in most of the considered $m_a$ range. 
 
\begin{figure}
\centering
\includegraphics[width=0.8\textwidth]{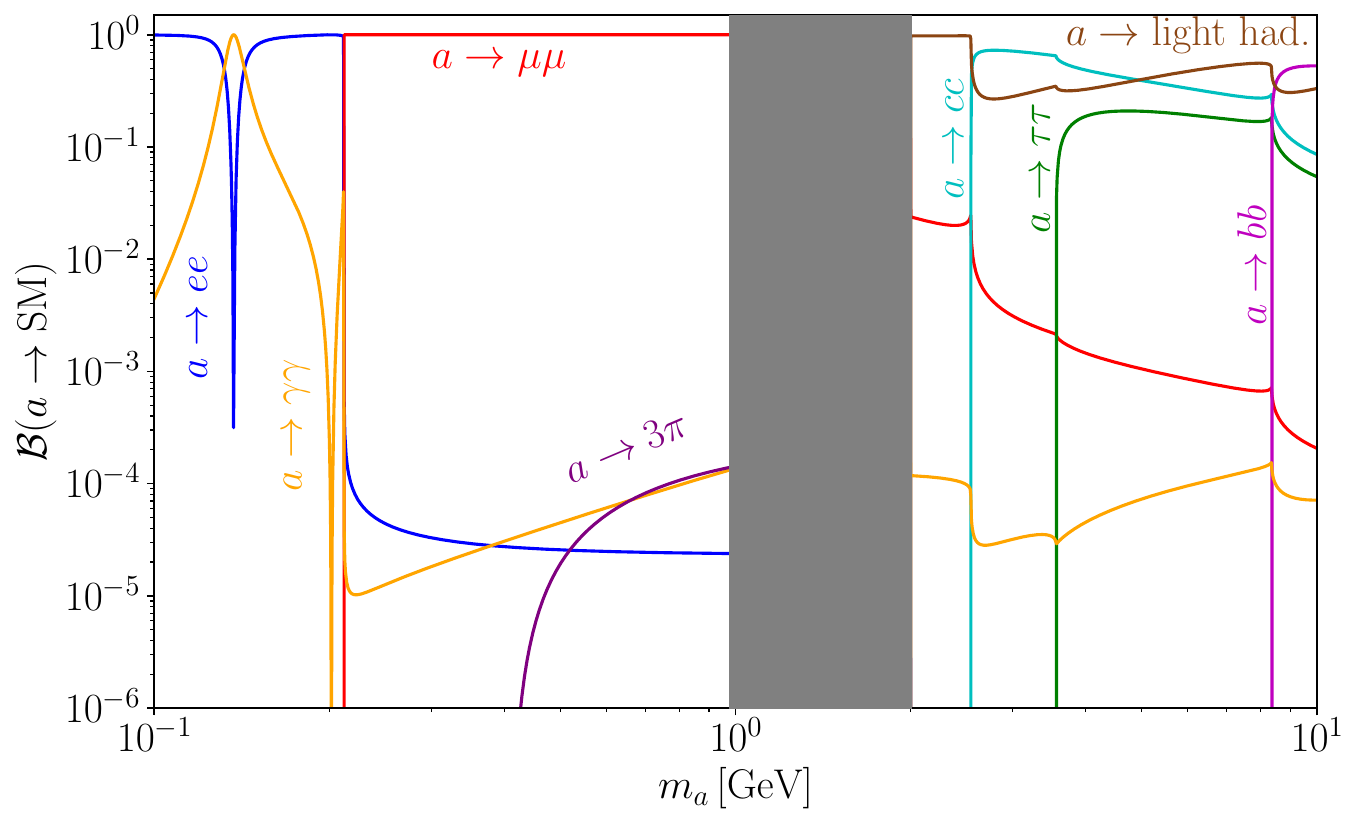}
\caption{\em Branching ratios of ALP decays into SM particles as a function of the ALP mass, setting all Wilson coefficients $c_{aXX}=1$ 
and $f_a = 1\,$TeV. } 
\label{fig:branchingratios}
\end{figure}

Finally, in a more general setup, one can consider the possibility in which the ALP is the portal to a light-dark sector. In the following, 
we will consider the simplest scenario described by the Lagrangian of  
Eq.~\eqref{eq:lagrangian3}, where a single light exotic fermion has been introduced with coupling $c_{a\chi\chi}$. 
Then, in all generality, the cross section for a specific ALP decay into the final state $X$ reads
\beq
\sigma\Big(e^+e^- \to V \to \gamma (a \to X)\Big) = \sigma_\mathrm{prod}(s) \, \mathcal{B}(a\to X) \, , 
\label{eq:visible_sigma}
\eeq 
where $\sigma_\mathrm{prod}(s)$ is the relevant production cross section of the quarkonia state (resonant, non-resonant 
or mixed) and $\mathcal{B}(a\to X)$ the branching fraction defined as
\bea
\label{eq:generalBR}
\mathcal{B}(a\to X) = \frac{\Gamma (a \to  X)}{\Gamma(a \to \mathrm{SM})+\Gamma(a \to \mathrm{DS})} \, .
\eea
Two simplified scenarios for the ALP decay are typically envisaged in the literature, dubbed respectively ``invisible" and ``visible". For the 
``visible'' ALP scenario we assume $c_{a \chi\chi} \approx 0$ and therefore $\mathcal{B}(a\to \mathrm{SM})\approx 1$. Then the 
relative strength of all SM contributions is the one already discussed and depicted in Fig.~\ref{fig:branchingratios}. Conversely, 
the ``invisible'' ALP scenario is obtained by imposing a large hierarchy between the ALP couplings to the SM and dark sector  
particles, i.e.~$c_{a\chi\chi}\gg c_{a \mathrm{SM}}$, which therefore leads to $\mathcal{B}(a\to \mathrm{SM})\approx 0$ and 
$\mathcal{B}(a\to\chi\chi)\approx 1$, within our simplified framework.

%%%%%%%%%%%%%%%%%%%%%%%%%%%%%%%%%%%%%%%%%%%%%%%%%%%%%%%%%%%%%%%
\section{Phenomenology of the ``invisible" and ``visible" ALP scenarios}
\label{sec:pheno}

In this section, we examine the phenomenological analysis of the $J/\Psi$ and $\Upsilon$ radiative quarkonia decays, 
addressing the ``invisible'' and ``visible'' ALP scenarios separately.

\subsection{The ``invisible'' ALP scenario}

In this case, the ALP decays invisibly inside the detector, with a very clean missing-energy signature and typically 
well-controlled SM backgrounds. In the specific case of radiative quarkonia decays, the main 
experimental signature is simply an isolated photon plus missing energy. This scenario is the most economical in terms of 
fitting parameters as it depends only on the ALP couplings $g_{a\gamma\gamma}$ and $g_{aQQ}$ (with $Q=c$ or $b$) 
through the production 
cross sections in Eqs.~\eqref{eq:non_resonant} and \eqref{eq:branching_ratio}. A detailed analysis of the invisible ALP decay 
at the $\Upsilon(nS)$ resonances can be found in Ref.~\cite{Merlo:2019anv}, while a more general study of hadronic and 
leptonic meson decays in invisible ALPs can be found in Refs.\cite{Guerrera:2021yss,Gallo:2021ame,Guerrera:2022ykl}. 
\begin{table}[!tb]
\centering
\begin{tabular}{c| c | c | c | c}
Exp. & Mass range [GeV] & Type of search & Resonance & ALP Decay   \\
\hline 
BaBar \cite{Babar:2008aby} & $0\, - \, 7.8\,$ & Mixed & $\Upsilon(3S)$ & Invisible  \\
BaBar \cite{BaBar:2010eww} & $0\, - \, 9.2 \,$ & Resonant &$\Upsilon(1S)$ & Invisible  \\
Belle \cite{Belle:2018pzt} & $0\, - \, 8.97\,$ & Resonant & $\Upsilon(1S)$ & Invisible   \\
Belle II \cite{Belle-II:2020jti} & $0.2 - 9.7$ & Non-resonant & $\Upsilon(4S)$ & $\gamma\gamma$ (recast) \\
BESIII \cite{BESIII:2020sdo} & $0\, -\, 1.2$ & Resonant & $J/\psi$ & Invisible\\
BESIII \cite{BESIII:2022rzz} & $0.165\, - \, 2.84$ & Resonant & $J/\psi$ & $\gamma\gamma$ (recast) \\
\end{tabular}
\caption{\em Summary of the experimental information used for the ``invisible" ALP scenario analysis. The searches are classified 
as resonant, non-resonant and mixed, according to the experimental setup.}
\label{tab:ExpInvisible}
\end{table}

\begin{figure}
\includegraphics[width=\textwidth]{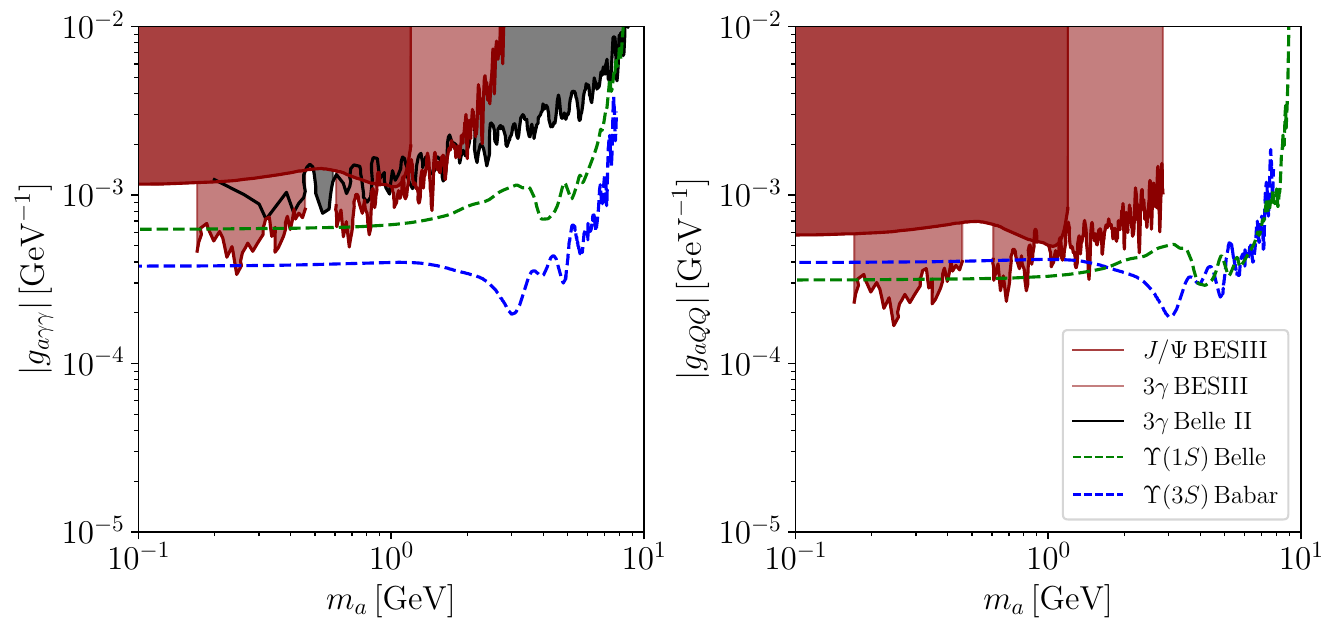}
\caption{\em Limits from the BESIII invisible (red) and tri-gamma recast (dark red) 
searches in the ``invisible'' ALP
scenario. The dashed blue, green and black lines represent respectively BaBar, Belle II invisible and the Belle II 
tri-gamma recast limits and are shown for comparison. 
In the right panel, $Q = c$ or $b$ respectively for 
Charm- or B-factories.} 
\label{fig:single_p_inv}
\end{figure}

A summary of the experimental searches considered in the ``invisible'' 
ALP analysis is provided in Tab.~\ref{tab:ExpInvisible}.
Although there are no new experimental data for the $\Upsilon(nS)$  
radiative decays into an ``invisible" ALP, with respect to the 
analysis provided in \cite{Merlo:2019anv}, BESIII has performed 
in \cite{BESIII:2020sdo}
a search for a $J/\Psi$ radiative decay 
through the process $\Psi(3686)\to \pi^+\pi^- J/\Psi$, followed by the decay $J/\Psi\to\gamma + \,\textrm{invisible}$, assuming 
in the final state an exotic neutral scalar decaying invisibly inside the detector. As the $J/\Psi$ is kinematically 
reconstructed from the original $\Psi(3686)$ decay process, one can assume the ALP being produced with the fully resonant 
cross section of Eq.~\eqref{eq:resonant}, thus acquiring a sensitivity to the $(c_{a\gamma\gamma},c_{acc})$ couplings for 
an ALP mass up to $m_a \lesssim 1.2\, $GeV. This analysis can complement B-factories limits on the ALP-photon coupling in the 
lower part of the ALP mass range, and provides new direct bounds on the ALP-charm interaction. In Fig.~\ref{fig:single_p_inv},
these new bounds are shown in red and compared with previous limits obtained in~\cite{Merlo:2019anv}, depicted as blue and 
green dashed lines. 

In addition to the invisible searches mentioned above, valuable information can also be obtained through the recast of recently 
published limits on the quarkonia ALP decay rates in the di-photon channel, $e^+e^-\to V\to\gamma\, (a\to\gamma\gamma)$ from 
BESIII \cite{BESIII:2022rzz} and Belle II~\cite{Belle-II:2020jti} collaborations, highlighting expectations from future 
``invisible" ALP searches at both these facilities. The BESIII search of Ref.~\cite{BESIII:2022rzz} looks for an ALP through 
the process $\Psi(3686)\to \pi^+\pi^- J/\Psi$, followed by the resonant decay $J/\Psi\to\gamma\,(a\to\gamma\gamma)$. In the 
BESIII analysis only the ALP-photon coupling is considered and consequently the ALP-SM branching ratio $\mathcal{B}(a\to
\gamma\gamma)=1$ is assumed. Using Eqs.~\eqref{eq:resonant}--\eqref{eq:branching_ratio}, and setting $c_{acc}=0$, 
from the BESIII limits on $\mathcal{B}(V\to\gamma a)$ one can derive bounds on the ALP-photon coupling. In the left panel 
of Fig.~\ref{fig:single_p_inv} our exclusion limits on $g_{a\gamma\gamma}$ are shown as light-red region, perfectly in 
agreement with the bounds presented in \cite{BESIII:2022rzz}.

Notice, however, that BESIII limits on $\mathcal{B}(V\to\gamma a)$ obtained from the tri-gamma channel can also be used to provide bounds on the foreseen sensitivities of the collaboration on the ALP-SM couplings in the ``invisible'' ALP scenario, to be compared with the $(g_{a\gamma\gamma}$, $g_{aQQ})$ limits derived from BESIII invisible, Belle and BaBar searches. From the two plots in Fig.~\ref{fig:single_p_inv} one can see that a future BESIII invisible ALP search would feature a sensitivity, up to the kinematically available mass region, fully comparable with B-factories limits. It should be stressed that such a limit on the $g_{acc}$ coupling, once provided, would represent the most stringent one in the considered ALP mass region, with B-factories yielding a comparable bound on the ALP-bottom coupling. 

\begin{figure*}[t!]
\centering
\includegraphics[width=0.9\textwidth]{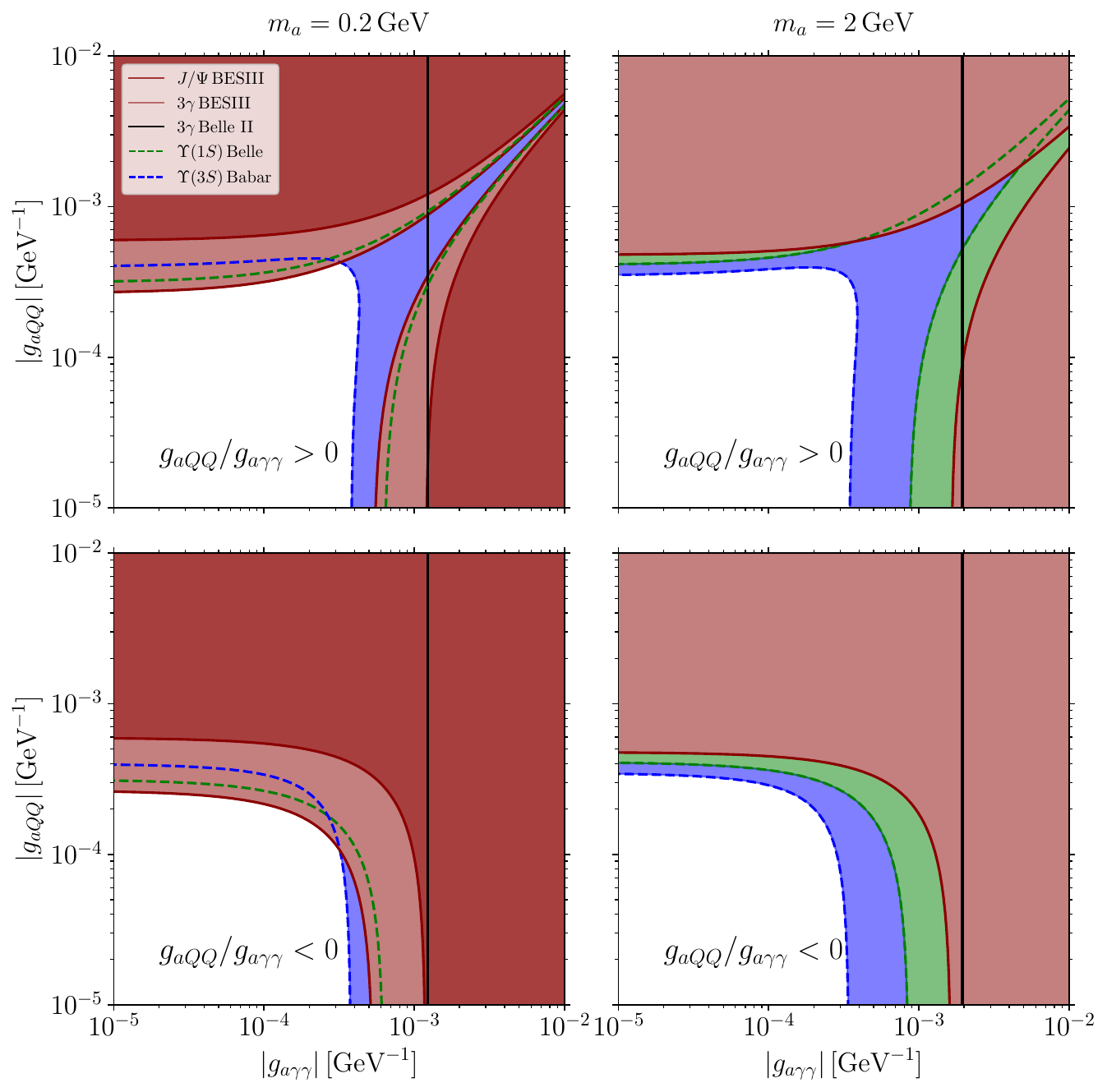}
\vspace{-0.2cm}
\caption{\em Excluded $(g_{a\gamma\gamma},g_{aQQ})$ parameter space in the ``invisible'' scenario, for two 
different values of the ALP mass $m_a=0.2\, $GeV (left) and $m_a=2\, $GeV (right) and for the two possible choices of the sign of the 
coupling ratio, $g_{aQQ}/g_{a\gamma\gamma}>0$ (upper) and $g_{aQQ}/g_{a\gamma\gamma}<0$ (lower). 
$Q = c$ or $b$ respectively for 
Charm- or B-factories. 
Dashed constraints are taken from 
\cite{Merlo:2019anv}. The dark-red area is from the BESIII search~\cite{BESIII:2022rzz}, while the continuous black line 
displays the
Belle II search~\cite{Belle-II:2020jti}.}
\label{fig:corr_inv}
\end{figure*}

On the same footing, Belle II has recently published in Ref.~\cite{Belle-II:2020jti} a similar analysis looking for an 
ALP through the process $e^+ e^- \to \Upsilon(4S) \to \gamma\,(a \to \gamma \gamma)$, thus providing bounds on the ALP-photon 
coupling, once $\mathcal{B}(a\to\gamma\gamma)=1$ is assumed. As explained before, the $\Upsilon(4S)$ resonance is so spread 
that the non-resonant ALP production cross section largely dominates and Eq.~\eqref{eq:non_resonant} has to be used in 
order to derive bounds in the ALP-photon coupling. Again, as before, the obtained bounds on the ALP non-resonant cross 
section can also be recast as bounds on the $g_{a\gamma\gamma}$ couplings in the ``invisible'' ALP scenario. The resulting 
exclusion limits are shown as a dark grey
region in the left panel of Fig.~\ref{fig:single_p_inv}, while clearly in this search 
no sensitivity can be obtained on the ALP-quark coupling.

Finally, we present in Fig.~\ref{fig:corr_inv} all the available information on the ``invisible'' ALP scenario from $J/\Psi$ 
and $\Upsilon$ radiative decays in the $(g_{a\gamma\gamma},g_{aQQ})$ parameter space, with $Q$ either $c$ or $b$ depending 
on the corresponding experiment 
and for two different reference values of the ALP mass, $m_a = 0.2$ (left) and $2\, $GeV(right). 
The dark and light red areas show the $(g_{a\gamma\gamma},\,g_{acc})$ exclusion regions from the invisible and tri-gamma 
(recast) BESIII analyses, while the blue and green dashed line indicate the bounds on $(g_{a\gamma\gamma},\,g_{abb})$ from 
BaBar and Belle $\Upsilon$ Invisible radiative decays (taken from Ref.~\cite{Merlo:2019anv}). The full black line shows, 
instead, the bound obtained from the Belle II tri-gamma (recast) analysis. In the two upper plots, one can observe 
the flat direction associated with resonant searches when $g_{aQQ}/g_{a\gamma\gamma}>0$ is chosen, absent instead when 
the opposite sign choice is performed (lower plots). This otherwise unconstrained direction in the $(g_{a\gamma\gamma},g_{aQQ})$ 
parameter space can be resolved only by means of the BaBar $\Upsilon(3S)$ mixed search (blue dashed line) or the 
Belle II non-resonant one (black continuous line).

As a final comment, notice that, at least in the sub-GeV mass range, BESIII has already the same potential sensitivity both on the ALP-photon and ALP-quark couplings than B-factories. 
This highlights the importance of undertaking a real ``invisible'' ALP 
search, updating the 2020 result of Ref.~\cite{BESIII:2020sdo} 
by using the present luminosity and extending the search up to $m_a=3\, $GeV.

\subsection{The ``visible'' ALP scenario}
%%%%%%%%%%%%%%%%%%%%%%%%%%%%%%%%%%%%%%%%%%%%%%%%%%%%%%%%%%%%%%%%%%%%

Let us consider now the case $\mathcal{B} (a\to \textrm{SM})\approx 1$, so that the ALP decays visibly into SM final 
states inside the detector. This scenario, dubbed ``visible'' ALP, 
opens up more experimental channels to look for, albeit 
at the cost of enlarging the ALP parameter space. 

\begin{table}[t]
\centering
\begin{tabular}{c| c | c | c | c}
Exp. & Mass range [GeV] & Type of search & Resonance & Decay   \\
\hline 
BaBar \cite{BaBar:2011kau} & $0.3\, - \, 7\,$ & Mixed & $\Upsilon(2S),\, \Upsilon(3S)$ & Hadrons  \\
BaBar \cite{BaBar:2012wey} & $0.212\, - \, 9.2\,$ & Resonant & $\Upsilon(1S)$ & $\mu^+\mu^-$  \\
BaBar \cite{BaBar:2015cce} & $4\, - \, 9.25\,$ & Resonant & $\Upsilon(1S)$ & $\overline{c} c$  \\
Belle \cite{Belle:2021rcl} & $2m_\ell - 9.2$ & Resonant & $\Upsilon(1S)$ & $\tau^+ \tau^-$, $\mu^+\mu^-$  \\
Belle II \cite{Belle-II:2020jti} & $0.2 - 9.7$ & Non-Resonant & $\Upsilon(4S)$ & $\gamma\gamma$  \\
BESIII \cite{BESIII:2021ges} & $0.212\, -\, 3$ & Mixed & $J/\psi$ & $\mu^+\mu^-$  \\
BESIII \cite{BESIII:2022rzz} & $0.165\, - \, 2.84$ & Resonant & $J/\psi$ & $\gamma\gamma$  \\
\end{tabular}
\caption{\em 
Summary of the experimental information used for the ``visible'' 
ALPscenario analysis. The searches are classified 
as resonant, non-resonant and mixed, according to the experimental setup. }
\label{tab:ExpVisible}
\end{table}

Several experiments have looked at different decay channels in the allowed ALP mass region. The most significant searches are 
summarised in Tab.~\ref{tab:ExpVisible}, 
where we have explicitly indicated also the ALP production mechanism, as described in
Sec.~\ref{sec:ALPprod}, and the associated decay mode. 
The oldest data come from the BaBar experiment, where searches in 
several decay channels have been released. For this analysis, we make use of the data from the $\Upsilon(1S)$ 
radiative decays \cite{BaBar:2012wey,BaBar:2013lkw,BaBar:2015cce} into a scalar particle, that subsequently decays visibly 
into muons and $c$-quarks. In these searches the  $\Upsilon(1S)$ is kinematically reconstructed from a higher quarkonium 
resonance decay and hence the contribution is resonant. In addition, we have also used the BaBar dataset of $\Upsilon(3S)$ and 
$\Upsilon(2S)$ decays into light and charm hadrons \cite{BaBar:2011kau}. 
Since the latter are untagged, 
the mixed ALP production cross section has been used in this case. Recall, however, that the ALP decay into light hadrons can 
only be approximately estimated (see discussion in Sec.~\ref{sec:ALPdecay}) and hence the bounds derived through these data 
should be interpreted with some caution, especially in the sub-GeV ALP mass range. Previous Belle visible searches 
\cite{Belle:2021rcl} with the ALP decaying into muons and taus are also considered in this work. The latest available analysis 
of radiative $\Upsilon$ decay into a visible ALP is the Belle II measurement of Ref.~\cite{Belle-II:2020jti} of the 
$\Upsilon(4S)$ decay into 3$\gamma$. As explained in Sec.~\ref{sec:ALPprod}, in this case, the non-resonant ALP production 
cross section has to be used, due to the very spread nature of this specific resonance. Newer data coming from BESIII 
complement B-factories searches with the ALP production via the radiative $J/\Psi$ decay and the subsequent ALP decay into 
muons \cite{BESIII:2021ges} and photons \cite{BESIII:2022rzz}. These data are important because they can provide additional 
direct information on the ALP-charm coupling, inaccessible at B-factories.  

Due to the large number of independent parameters entering in the ``visible" ALP decay scenario, in the following, we will 
present our phenomenological analysis in two simplified frameworks, 
considering first universal ALP-fermion couplings and then employing the 
two benchmark ALP models introduced in Sec.~\ref{sec:UVmodels}.

\subsubsection*{Universal ALP-fermion coupling}

As a first simplified scenario, we study how available experimental data from radiative quarkonia decays can constraint 
the ALP parameter space by assuming a universal ALP-fermion coupling, denoted hereafter as $c_{aff}$. This 
universal ALP-fermion coupling scenario can be theoretically motivated by generating all the ALP-fermion couplings 
exclusively via the ALP-Higgs operator $(\partial_\mu a) H^\dagger 
\overset{\leftrightarrow}{D^\mu} H$, see for instance Ref.~\cite{Georgi:1986df}. 
We neglect here, for simplicity, the small differences in the individual ALP-fermion couplings induced by running effects. 

\begin{figure}[t]
\centering
\includegraphics[width=0.8\textwidth]{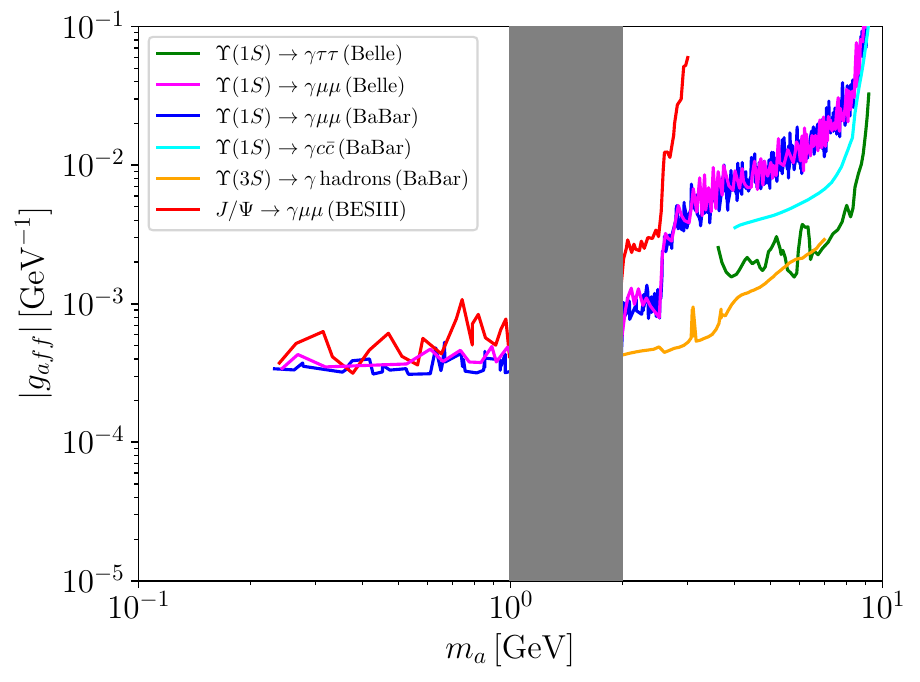}
\caption{\em  BESIII, Belle(II) and BaBar 90\% exclusion bounds obtained from visible ALP decay searches, in the universal 
ALP-fermion scenario and assuming vanishing tree-level ALP-gauge anomalous couplings. } 
\label{fig:visible_1coupling}
\end{figure}

In Fig.~\ref{fig:visible_1coupling} the bounds on the universal ALP-fermion coupling, $g_{aff}\equiv c_{aff}/f_a$,  
obtained from radiative quarkonia decays into a visible ALP are shown, assuming vanishing tree-level ALP-gauge anomalous 
Wilson coefficients, $c_{agg}=c_{a\gamma\gamma}=0$. However, for consistency we are including the fermion loop-induced 
contributions, proportional to $c_{aff}$, as discussed in Sec.~\ref{sec:ALPdecay}. 
In Fig.~\ref{fig:visible_1coupling} one can see that for $m_a \lesssim 1\, $GeV muon searches at Charm and B-factories give very 
similar results, $g_{aff} \lesssim 3\times 10^{-3}\, $GeV$^{-1}$ at 90\% CL. No relevant constraints can be obtained in the 
ALP low-mass region from ALP hadronic decays as the ALP decay width into light-hadrons of Eq.~\eqref{eq:lighth} is almost 
vanishing due to an ``accidental'' cancellation that takes place in the ALP-fermion universal scenario, i.e.~$c_{auu}=
c_{add}$, leading to $c_{a\pi}\approx 0$, see Eq.~\eqref{eq:capi}.  
Conversely, at higher ALP masses, $m_a \gtrsim 2\, $GeV, the strongest bounds on the universal ALP-fermion coupling come from BaBar 
searches in ALP-hadron (yellow line) and ALP-tau (green line) decays when kinematically allowed, with all the other channels 
providing bounds 5-10 times weaker and with BESIII data still subleading compared to the same BaBar/Belle channel. 

Bounds on $g_{a\gamma\gamma}$ for the ``visible" ALP scenario, assuming a vanishing universal fermion coupling, $g_{aff}=0$, 
are not shown here as they coincide with the excluded light-pink and grey regions depicted in the left plot 
of  Fig.~\ref{fig:single_p_inv}, obtained from the BESIII and Belle II 3$\gamma$ searches of Ref.~\cite{BESIII:2022rzz,
Belle-II:2020jti}.

\begin{figure}[t!]
\includegraphics[width=1\textwidth]{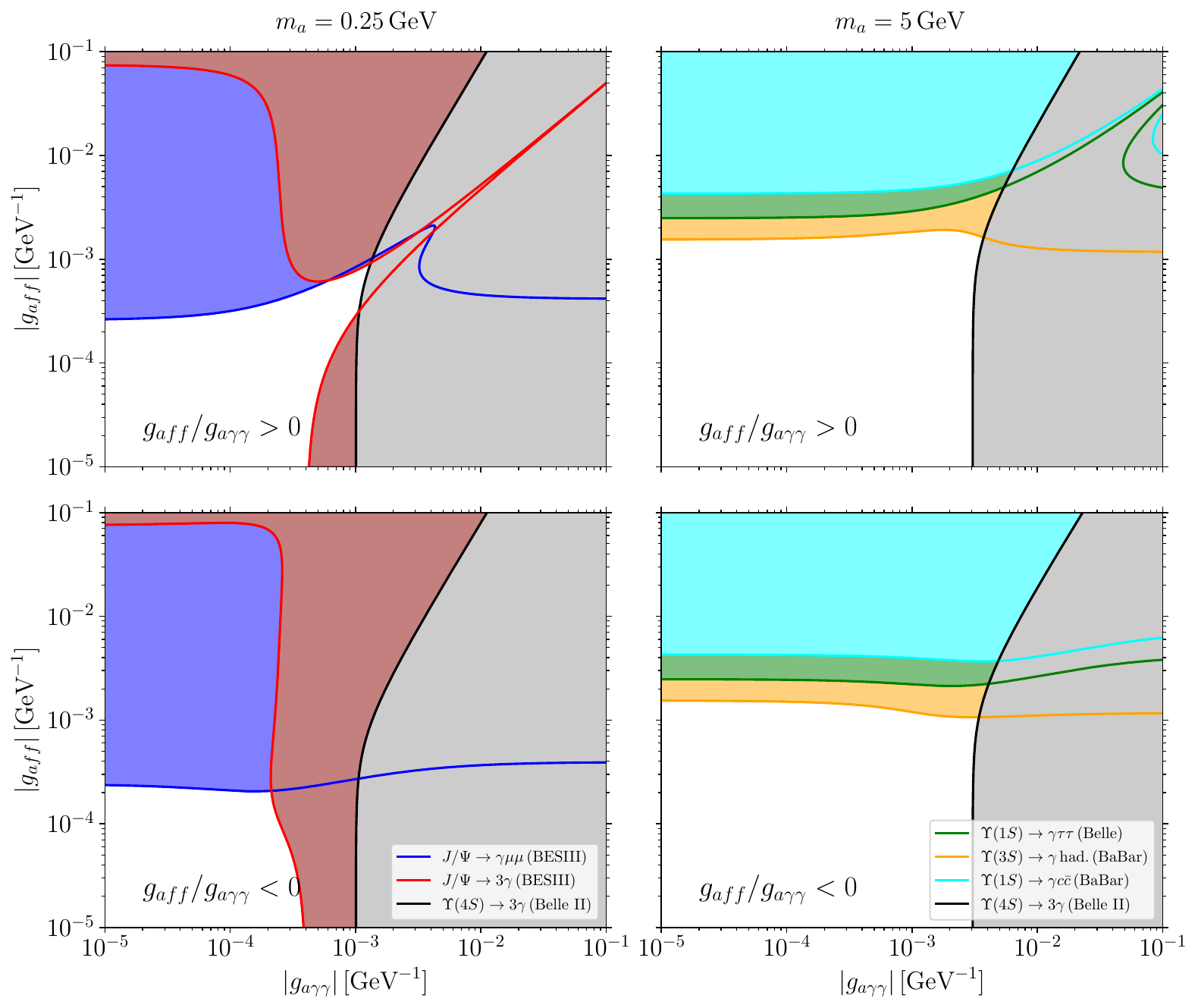}
\caption{\em Excluded parameter space for the two benchmark values $m_a=0.25\,$GeV (\textit{left}) and $m_a=5\,$GeV (\textit{right}) 
and for the two possible choices of the sign of the 
coupling ratio, $g_{aff}/g_{a\gamma\gamma}>0$ (upper) and $g_{aff}/g_{a\gamma\gamma}<0$ (lower). 
The blue line indicates the BESIII search of an ALP decaying into two muons, which largely overlaps with the BaBar and Belle searches.}
\label{fig:visible_2couplings}
\end{figure}

Finally in Fig.~\ref{fig:visible_2couplings} the combined bounds on $(g_{a\gamma\gamma},g_{aff})$ from the ``visible" ALP 
searches are summarised, assuming a vanishing tree-level ALP-gluon coupling, $g_{agg}=0$. In the left plots, the 
excluded parameter space is shown for a value of the ALP mass well inside the region of validity of chiral perturbation 
theory, i.e.~$m_a=0.25\, $GeV. In these plots, for the sake of simplicity, we choose to display the BESIII muons data alone, 
as all muons constraints  practically overlap, as learned from Fig.~\ref{fig:visible_1coupling}. In the upper/lower 
plots the bounds are shown for the two alternative sign choices, $g_{aff}/g_{a\gamma\gamma} \lessgtr 0$.
In the upper-left plot, one can observe that the typical flat direction present in all resonant channels (red line) 
is resolved by the mixed BESIII muon channel and by the non-resonant 3$\gamma$ Belle II search at the $\Upsilon(4S)$ 
resonance. Therefore, once again the complementarity of searches with a different ALP production mechanism appears evident. 
Clearly, no flat direction appears when the negative sign is chosen.  

In the right plots of Fig.~\ref{fig:visible_2couplings} similar bounds are shown but for an ALP mass well inside the perturbative QCD 
validity range, i.e.~$m_a=5\, $GeV. For this large ALP mass, only data from B-factories are available. The resonant data from tau 
and c-quark ALP decays give constraints slightly weaker than 
the mixed ALP-hadrons decay channel. However, all these 
experiments give relevant bounds only on the ALP-fermion coupling while the non-resonant Belle II 3$\gamma$ search becomes 
fundamental in closing the two-dimensional parameter space. In this case, the tau and hadronic searches are the 
most relevant ones, setting a robust bound on the universal coupling $g_{aff}\lesssim 1-2\times 10^{-3}\, $GeV$^{-1}$. 
Finally, it can be seen that bounds on the fermion coupling are an order of magnitude better in the low-mass scenario 
than in the heavier ALP case. This fact can already be noticed from looking at the individual coupling bound of 
Fig.~\ref{fig:visible_1coupling}, and similarly for the photon coupling in Fig.~\ref{fig:single_p_inv}, and it is associated to a phase space suppression at higher ALP masses.

\subsubsection*{Benchmark ALP models}

Another possibility, in order 
to reduce the number of free parameters in the 
``visible'' ALP scenario, is to consider the UV-complete ALP models (previously introduce in Sec.~\ref{sec:UVmodels})
which provide non-trivial correlations among the Wilson coefficient, thus considerably 
reducing the ALP parameter space. 
Notice that in the effective ALP Lagrangian of Eq.~(\ref{eq:lagrangian2}) we have assumed that top-quark (and the 
heavy weak bosons) contributions were already matched into the low-energy Wilson coefficients. However, when considering 
specific UV-complete models these contributions need to be included explicitly. To do so, we implement in our numerical 
analysis the analytical results of Ref.~\cite{Bauer:2021mvw} and take into account the effects of running from the high-energy 
scale $f_a$ ($1-10\,$TeV in our benchmark cases) and integrating out the top at the electroweak scale. This procedure can account up 
to a $10\%$ modification of the UV couplings defined in Eq.~\eqref{eq:couplDFSZQED} and \eqref{eq:couplKSVZ}. 
On the other hand, the weak boson contributions are safely negligible \cite{Bauer:2021mvw}. 

\begin{figure}[t!]
\centering
\includegraphics[width=0.49\textwidth]{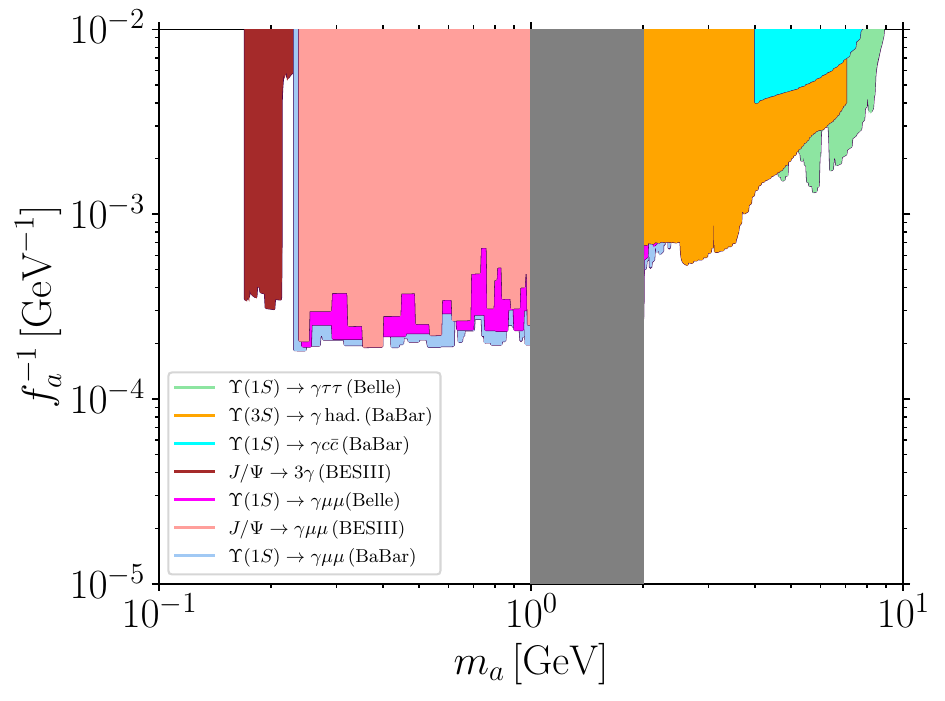} 
\includegraphics[width=0.49\textwidth]{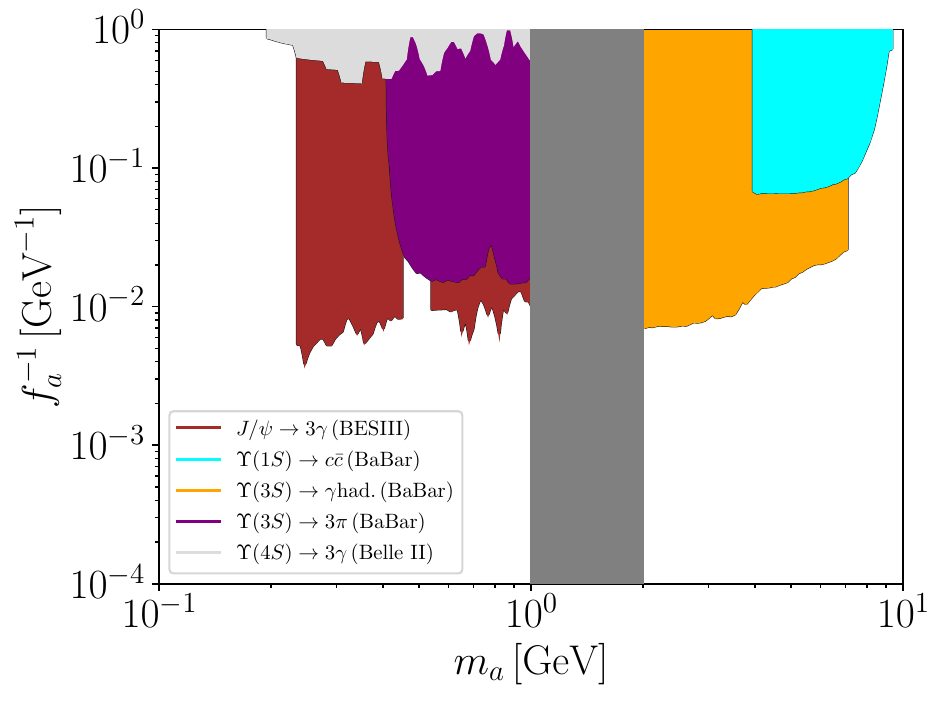}
\caption{\em Bounds on $1/f_a$ as function of the ALP mass, $m_a$, 
for the two benchmark ALP models 
considered in the text: DFSZ-QED ALP (left) with $\tan\beta=1$
and KSVZ ALP (right).}
\label{fig:BenchmarkModels}
\end{figure}

The first model we have considered is the DFSZ-QED ALP, which is anomalous under electromagnetism but not under QCD. It is not 
completely fermion universal as all ALP-quarks couplings are the same but different from the ALP-lepton ones. 
In the DFSZ-QED model, there are three independent parameters, $m_a,\, f_a$ and  $\tan\beta$. 
In the following, we will set for simplicity $\tan\beta=1$, as in this point both couplings to leptons and quarks have 
the same magnitude. For this benchmark point, we plot in the left panel of Fig.~\ref{fig:BenchmarkModels} the excluded 
$1/f_a$ region as a function of the ALP mass $m_a$. Notice that photon searches are dominant only when the ALP decay 
into muons is not kinetically allowed. The BESIII bound is sizeable because the production of the ALP is made via the $c$ coupling, while the 3$\gamma$ Belle II search is 
of non-resonant type and hence is only sensitive to the photon coupling suppressed by $\alpha_{em}^2$ in our conventions of 
Eq.~\eqref{eq:lagrangian2}. When the ALP decay into muons is allowed this channel becomes the most stringent one below $1\,$GeV, 
with all experiments having similar sensitivities. At higher masses, hadronic and tau decays set the most stringent bounds, 
while $c\bar c$ exclusive decays are less relevant. As all fermion couplings are generated at tree level in this model the 
bounds are similar to those of the universal scenario shown in Fig.~\ref{fig:visible_1coupling}, ranging from $f_a\sim 10\,$TeV for lower masses and $f_a\sim 1\,$TeV in the heavier mass region.

In the second ALP model considered, the KSVZ ALP, all couplings are generated via the $c_{agg}$ 
anomalous term: including one-loop-induced quark couplings (see Ref.~\cite{Arias-Aragon:2022iwl}), and a photon coupling 
via mixing with the pion. In this case, one has only two independent parameters $\left(m_a,\,f_a\right)$, and expects 
in general weaker constraints on the scale $f_a$. We do not include in this study the two-loop ALP-lepton and ALP-photon 
contributions, as they are safely negligible. The obtained bounds on $1/f_a$ as a function of the ALP mass are shown in 
the right panel of Fig.~\ref{fig:BenchmarkModels}. For low values of $m_a$ the 3$\gamma$ BESIII search is the most relevant, 
together with the $3\pi$ channel, once kinematically allowed. Limits from Belle II are very weak, as the production is only 
via the photon coupling and has therefore an extra $\alpha_{\textrm{em}}$ suppression. Similarly to the DFSZ-QED ALP benchmark, 
at larger masses, the strongest bound is set by the hadronic decays, while the limits from exclusive $c-$quark searches are still an 
order of magnitude weaker.

\section{Conclusions}
\label{sec:concl}

In this paper we have revisited the ALP production in $\Upsilon$ and $J/\psi$ radiative decays. The ALP production can 
proceed through three distinct and complementary experimental setups, dubbed as non-resonant, resonant and mixed.  
Depending on the specific production mechanism, they enable to access different combinations of ALP-SM particle couplings. 
We have then considered both the ``invisible'' and ``visible'' ALP decay scenarios, depending respectively on whether the 
ALP escapes detection (e.g.~by decaying into an unspecified dark sector) or decays into SM final states. 

For the ``invisible'' ALP scenario, we have updated the analysis presented in Ref.~\cite{Merlo:2019anv} by adding the constraints 
on the $(g_{a\gamma\gamma}, g_{acc})$ parameter space from the $J/\psi\to\gamma+E_{mis}$ BESIII analysis \cite{BESIII:2020sdo}. 
In addition, we presented here the projected sensitivity on the $(g_{a\gamma\gamma},g_{acc})$ and $(g_{a\gamma\gamma},g_{abb})$ 
parameter space of future ``invisible'' ALP searches, obtained by recasting the recent BESIII $J/\Psi\to\gamma a\, 
(a\to\gamma\gamma)$ search \cite{BESIII:2022rzz} along with the constraints from the analogous Belle II $\Upsilon(4S)\to\gamma 
a\,(a\to \gamma\gamma)$ channel \cite{Belle-II:2020jti}. ALP-photon and ALP-$Q$ couplings, with $Q=c,b$, larger than roughly 
$10^{-3}\, $GeV$^{-1}$ are already excluded by present data for ALP masses in the $m_a \in [0.1,10] \, $GeV range. Assuming the 
already achieved BESIII and Belle II luminosities, a factor of five improvement in the above limits is expected when the new 
analyses in the mono-$\gamma$ plus missing-energy channel will become available. 

In the case of a ``visible'' ALP decay, two $m_a$ regions have been separately considered, respectively $m_a \lesssim 1\, $GeV 
and $m_a \gtrsim 2\, $GeV. This is because the ALP decay width, $\Gamma(a\to \text{SM})$, turns out to be difficult to estimate in the intermediate 1-2 GeV region, where neither chiral perturbation theory nor perturbative QCD can be applied. 
The landscape of currently available searches for visible ALP decay channels from radiative quarkonia decays has been 
then presented. To streamline the analysis of the “visible” ALP scenario, we have introduced additional theoretical 
assumptions, such as ``universal ALP-fermion couplings'' or specific ``benchmark ALP models'', aimed to reduce the number 
of independent parameters involved. As a general feature, we noticed that the strongest constraints on the ALP-SM couplings 
in the sub-GeV $m_a$ range come from the BESIII and Belle II muon and 3$\gamma$ channels, with an almost equivalent 
sensitivity on $f_a \gtrsim 10^3-10^4\, $GeV. For the multi-GeV $m_a$ case the bounds are typically one order of magnitude 
weaker than for the lower ALP mass range, with the strongest constraints coming from hadronic channels at B-factories.

From the analysis presented in this paper clearly emerges the relevance of quarkonia decays in constraining the ALP parameter space. Further significant improvements on the ALP-fermion flavour-conserving couplings can be envisaged when/if 
future experimental searches at B- and Charm-factories will be available. In particular, BESIII searches in the light hadrons 
channel could be extremely useful to further constrain the ALP coupling to pions. 
In the multi-GeV ALP range, instead, Belle II radiative decays of $\Upsilon(4S)$ into both 
invisible and visible channels, 
besides the already available 3$\gamma$ channel, would offer a complementary 
set of information thanks to the underlying non-resonant ALP production mechanism. Finally, quarkonia decays are also very promising for constraining ALP-fermion flavour-violating couplings, which were not addressed in this study but will be the focus of a future work.

\section*{Acknowledgments}

The authors would like to thank Stefano Lacaprara, Michael De Nuccio and Giovanni Verza for useful suggestions and 
discussions. This work received funding from the European Union's Horizon 2020 research and innovation programme under 
the Marie Sk\l{}odowska-Curie grant agreements n. 860881 -- HIDDeN, n.~101086085 -- ASYMMETRY and by the INFN Iniziative 
Specifica APINE. This work was also partially supported by the Italian MUR Departments of Excellence grant 2023-2027 
``Quantum Frontiers".
The work of LDL is supported by 
the European Union -- NextGenerationEU and by the University of Padua under the 2021 STARS Grants@Unipd programme (CPV-Axion -- Discovering the CP-violating axion) as well as by
the European Union -- Next Generation EU and by the Italian Ministry of University and 
Research (MUR) via the PRIN 2022 project n.~2022K4B58X -- AxionOrigins. 

\appendix

\section{Exact diagonalization of the ALP-pion mixing}
\label{app:diagonalization}

When the mass of the ALP is below the GeV scale a mixing between the pion and the ALP arises from the chiral Lagrangian, 
see e.g.~Refs.~\cite{Bauer:2017ris,Bauer:2020jbp,DiLuzio:2022tbb}. 
In this Appendix we report for completeness the exact diagonalization of the ALP-pion system, which is usually treated perturbatively.  
This is especially useful for the transition region $m_a \approx m_\pi$, where the perturbative diagonalization breaks down. 

\begin{figure}[t!]
    \centering
\includegraphics[width=0.7\textwidth]{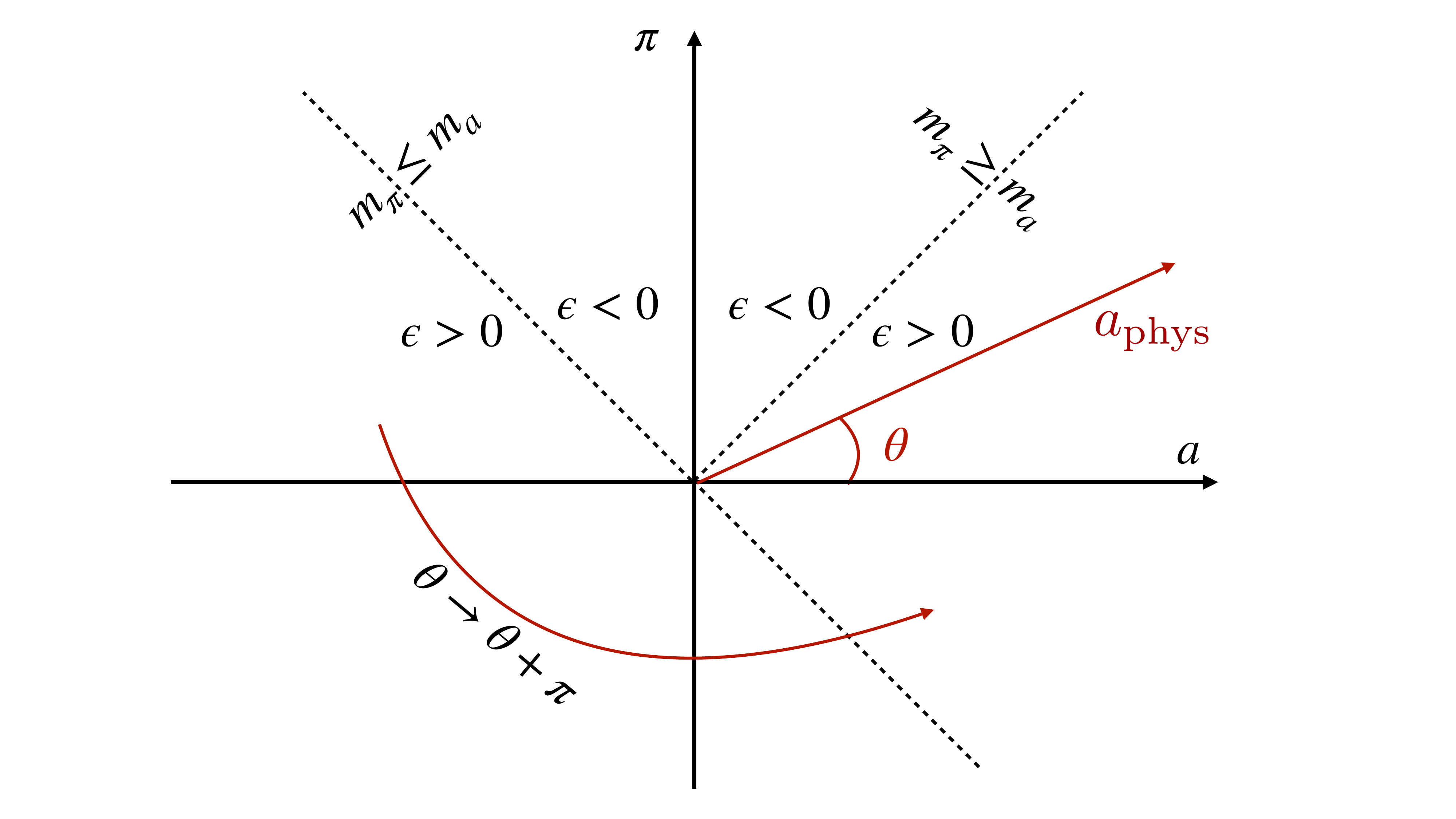}
    \caption{\em Sketch of the different regions 
    for the angle $\theta$, 
    depending on the ALP-pion mass difference and the sign of $\epsilon$.
    }
    \label{fig:sketch}
\end{figure}

In the notation of Ref.~\cite{DiLuzio:2022tbb}, 
to which we refer for the derivation of the ALP-pion chiral Lagrangian, the ALP-pion mixing stems from the quadratic terms
\beq
\mathcal{L}_{a\pi}
= \frac{1}{2}
\begin{pmatrix} \partial_\mu a & \partial_\mu \pi^0 \end{pmatrix} \mathcal{K}_{\textrm{LO}}\begin{pmatrix} \partial^\mu a \\ \partial^\mu \pi^0 \end{pmatrix} - \frac{1}{2}\begin{pmatrix} a &  \pi^0 \end{pmatrix} \mathcal{M}^2_{\textrm{LO}}\begin{pmatrix} a \\  \pi^0 \end{pmatrix} \, ,
\end{equation}
with 

\begin{equation}
 \mathcal{K}_{\textrm{LO}}=\begin{pmatrix}
        1 & \epsilon \\
        \epsilon & 1
    \end{pmatrix}\quad \textrm{and} \quad \mathcal{M}^2_{\textrm{LO}}=\begin{pmatrix}
        m_a^2 & 0 \\
        0 & m_\pi^2
    \end{pmatrix}\, . 
\end{equation}
Here, $m_a$ and $m_\pi$ are the leading order ALP and pion mass parameters,  
while 
$\epsilon$ is defined as
\beq
 \epsilon =-\frac{1}{2}\frac{f_\pi}{f_a} \left(2 c_{agg} \frac{m_d-m_u}{m_u+m_d} + c_{auu} -c_{add}\right)
 = \frac{1}{2} \frac{f_\pi}{f_a}c_{a\pi} 
 \, . 
\eeq
The first step is to diagonalize and then re-scale the kinetic term in order to make it canonical. This is obtained via the matrix
\beq
    W_K = \frac{1}{\sqrt{2}}\begin{pmatrix} \frac{-1}{\sqrt{1-\epsilon}} & \frac{1}{\sqrt{1+\epsilon}}\\
    \frac{1}{\sqrt{1-\epsilon}} & \frac{1}{\sqrt{1+\epsilon}}    
    \end{pmatrix} \, , 
\eeq
where fields are understood to be multiplied by the inverse of $W_K$.
After applying this transformation, the mass matrix becomes non-diagonal
\beq
    W_K^T \mathcal{M}^2_{\textrm{LO}} \, W_K = \frac{1}{2}\begin{pmatrix}
        \frac{m_\pi^2+m_a^2}{1-\epsilon} & \frac{m_\pi^2-m_a^2 }{\sqrt{1-\epsilon^2}}\\
        \frac{m_\pi^2-m_a^2 }{\sqrt{1-\epsilon^2}} &\frac{m_\pi^2+m_a^2 }{1+\epsilon}
    \end{pmatrix} \, .
\eeq
This $2\times 2$ symmetric matrix can be now 
diagonalized using an 
orthogonal transformation, that is parametrized in terms of an angle $\theta$
\begin{equation}
    U_\theta = \begin{pmatrix}
        \cos{\theta} & -\sin{\theta} \\
        \sin{\theta} & \cos{\theta}
    \end{pmatrix}\, , 
\end{equation}
with
\beq
\label{eq:tan2theta}
    \tan{2\theta}=\frac{m_\pi^2-m_a^2}{m_\pi^2+m_a^2}\frac{1-\epsilon^2}{\epsilon}\, ,  
\eeq
and eigenvalues
\begin{align}
\label{eq:m1}
m_1^2 &= \frac{m_a^2+m_\pi^2+\sqrt{(m_a^2-m_\pi^2)^2+4\epsilon m_a^2m_\pi^2}}{2(1-\epsilon^2)} \, , \\  
\label{eq:m2}
m_2^2 &= \frac{m_a^2+m_\pi^2-\sqrt{(m_a^2-m_\pi^2)^2+4\epsilon m_a^2m_\pi^2}}{2(1-\epsilon^2)}  \, .
\end{align}
Note that shifting $\theta \to \theta + \pi$ 
one simply multiplies $U_{\theta}$ by an overall 
minus, which does not affect the diagonalization. 
Hence, it is possible to set the domain of the angle 
in $0 \leq \theta < \pi$. 
However, since $\tan{2\theta} = \tan{(2\theta + \pi)}$, 
Eq.~(\ref{eq:tan2theta}) does not allow one to distinguish between the intervals 
$0 \leq \theta \leq \pi/2$ and 
$\pi/2 \leq \theta < \pi$. 
To this end, it is useful to 
consider
\begin{align}
\label{eq:sin2theta}
    \sin 2\theta &= \frac{(m_\pi^2-m_a^2)\sqrt{1-\epsilon^2}}{\sqrt{(m_a^2-m_\pi^2)^2+4\epsilon m_a^2m_\pi^2}} \, , \\
\label{eq:cos2theta}    
    \cos 2\theta &= \epsilon\frac{m_\pi^2+m_a^2}{\sqrt{(m_a^2-m_\pi^2)^2+4\epsilon m_a^2m_\pi^2}} \, , 
\end{align}
which allow one to determine the quadrant in which the angle lies. For instance, from Eq.~(\ref{eq:sin2theta}) 
we have that according to the sign of $m_\pi^2-m_a^2$: 
$0 \leq \theta \leq \pi/2$ for $m_\pi^2 \geq m_a^2$ 
and 
$\pi/2 \leq \theta < \pi$ for $m_\pi^2 \leq m_a^2$.
On the other hand, from Eq.~(\ref{eq:cos2theta}) we 
have that according to the sign of $\epsilon$: 
$-\pi/4 \leq \theta \leq \pi/4$ for $\epsilon \geq 0$ 
and 
$\pi/4 \leq \theta \leq 3\pi/4$ for $\epsilon \leq 0$.  
Note that the region $-\pi/4 \leq \theta \leq 0$ 
can be equivalently mapped into 
$3\pi/4 \leq \theta \leq \pi$, 
thanks to the shift $\theta \to \theta + \pi$. 
A sketch of the different regions 
for the diagonalization angle 
is provided 
in Fig.~\ref{fig:sketch}. 

In order to implement the transformation that diagonalizes the mass matrix on the ALP and pion fields, it is actually more convenient to provide an expression directly for the trigonometric functions:
\begin{align}
\label{eq:sintheta}
    \sin\theta=\pm\frac{1}{\sqrt{2}}\left(1- \epsilon \frac{m_a^2+m_\pi^2}{\sqrt{(m_a^2-m_\pi^2)^2+4 \epsilon^2 m_a^2 m_\pi^2 }}\right)^{1/2}\, , \\
\label{eq:costheta}    
    \cos\theta=\pm\frac{1}{\sqrt{2}}\left(1+ \epsilon \frac{m_a^2+m_\pi^2}{\sqrt{(m_a^2-m_\pi^2)^2+4 \epsilon^2 m_a^2 m_\pi^2 }}\right)^{1/2} \, .
\end{align}
Here, the signs should be properly chosen according to 
the discussion above (cf.~also Fig.~\ref{fig:sketch}). 
Taking for instance $\epsilon>0$
and $m_\pi>m_a$ we have $0<\theta<\pi/4$, and hence $\cos\theta>\sin\theta$ are both positive. 

We then conclude that the outer signs 
in Eqs.~(\ref{eq:sintheta})-(\ref{eq:costheta}) 
need to be both 
positive. 
On the other hand, for $\epsilon>0$
and $m_\pi<m_a$, we have 
$3\pi/4<\theta<\pi$, and hence we need to choose 
plus for the sine in Eq.~(\ref{eq:sintheta}) and 
minus for the cosine in Eq.~(\ref{eq:costheta}). 
Note that the sign of $\epsilon$ will determine whether the sine is larger or smaller than the cosine. 
It is also useful to note that it is equivalent to diagonalise doing a $\theta+\pi$ rotation. This is helpful if one wants to display a dependence 
of a certain observable with respect to $m_a$, as if we diagonalise in the case $\epsilon>0$ and cross $m_a=m_\pi$ then one needs to switch the sign of the cosine in a 
non-continuous way. 

Then the full diagonalization 
matrix acting on the physical 
mass eigenstates
can be defined as
\begin{equation}
    \begin{pmatrix}
        a \\
        \pi
    \end{pmatrix} \equiv U_{a\pi} P_{\textrm{phys}}=
    \frac{1}{\sqrt{2}}\begin{pmatrix}
     \frac{-c_\theta}{\sqrt{1-\epsilon}}+\frac{s_\theta}{\sqrt{1+\epsilon}} & \frac{s_\theta}{\sqrt{1-\epsilon}}+\frac{c_\theta}{\sqrt{1+\epsilon}}\\
     \frac{c_\theta}{\sqrt{1-\epsilon}}+\frac{s_\theta}{\sqrt{1+\epsilon}} & \frac{-s_\theta}{\sqrt{1-\epsilon}}+\frac{c_\theta}{\sqrt{1+\epsilon}}
    \end{pmatrix} \begin{pmatrix}
        P^\textrm{phys}_1 \\
        P^{\textrm{phys}}_2
    \end{pmatrix} \, , 
\end{equation}
where we have simplified the notation 
by defining $c_\theta\equiv\cos\theta$ and $s_\theta\equiv\sin\theta$, and $P^\textrm{phys}_1 $ and $P^\textrm{phys}_2$ are the physical states associated to the eigenvalues $m_1$ and $m_2$. We will identify these states with what we call as the ``physical'' ALP 
and pion fields later on, as this requires one last argument. 

\begin{figure}
    \centering
\includegraphics[width=0.5\textwidth]{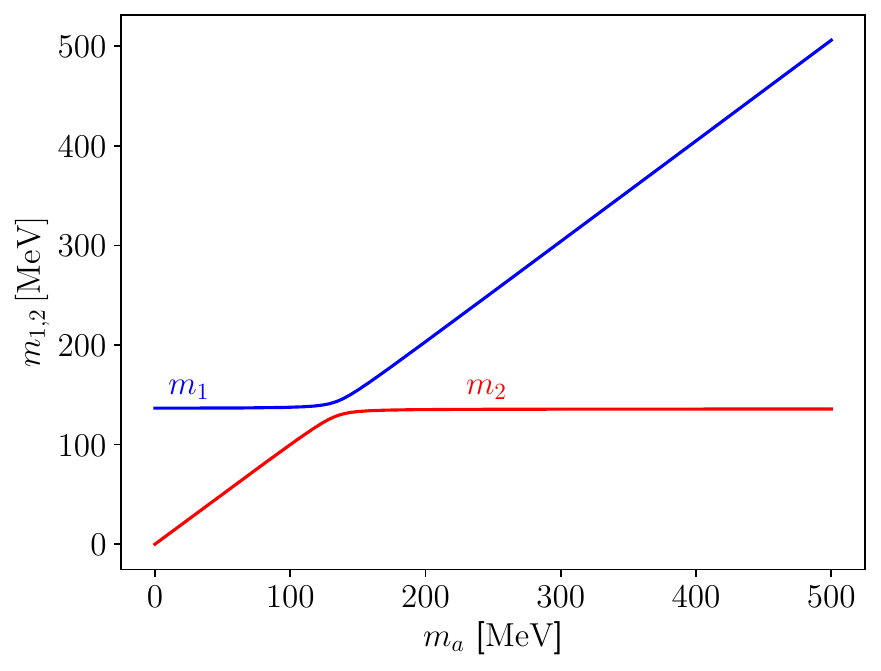}
\includegraphics[width=0.49\textwidth]{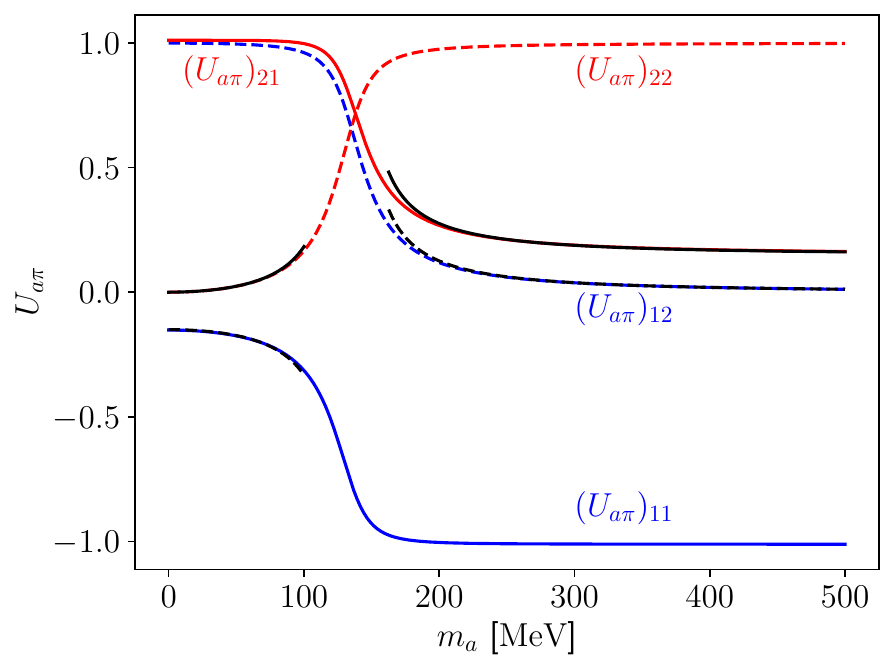}
    \caption{
    \em Eigenvalues (left panel) and mixing elements between the states $a$ and $\pi$ (right panel) 
    as a function of $m_a$, 
    for $\epsilon=0.1$ and $m_\pi=135\,$MeV. Blue (red) colour refers to the elements related to the 1st (2nd) physical state, while black lines represent 
    approximate formulae which break down for 
    $m_a \approx m_\pi$.}
    \label{fig:rotation}
\end{figure}

The eigenvalues are a combination of the original mass parameters (cf.~Eqs.~(\ref{eq:m1})-(\ref{eq:m2})). 
In the left panel of Fig.~\ref{fig:rotation} 
we fix $m_\pi = 135\,$MeV and plot 
the eigenvalues as a 
function of $m_a$. Note that there is always one eigenvalue that remains closer to $m_\pi$, while the other one follows $m_a$. 
For this reason, we associate the physical states 
$\pi_\textrm{phys}$ and $a_\textrm{phys}$ to those 
states whose mass is closer to $m_\pi$ and 
$m_a$, respectively. 
From Fig.~\ref{fig:rotation} we actually see that 
this identification has to be flipped when 
$m_a$ crosses through $m_\pi$. 
Hence, $\pi_\textrm{phys} = P_1^\textrm{phys}$ for $m_a < m_\pi$, while $\pi_\textrm{phys} = P_2^\textrm{phys}$ 
for $m_a > m_\pi$. The mixing elements need to be changed accordingly, for instance, the pion state is composed of 
\begin{align}
    \pi = (U_{a\pi})_{21} \pi_\textrm{phys}+ (U_{a\pi})_{22} a_\textrm{phys} \quad \textrm{for \ } m_a < m_\pi  \, , \\
    \pi = (U_{a\pi})_{21} a_\textrm{phys}+ (U_{a\pi})_{22} \pi_\textrm{phys} \quad \textrm{for \ } m_a > m_\pi \, . 
\end{align}
An important consequence of this identification 
concerns the ALP coupling to photons, 
provided in Eq.~(\ref{eq:caeffgaga}), 
which results in
\begin{equation}
    c_{a\gamma\gamma}^{\textrm{eff}}=c_{a\gamma\gamma}-1.92 c_{agg}+\frac{f_a}{f_\pi} U_{a\pi} \, , 
    \label{eq:photon_effective}
\end{equation}
where the ALP-pion mixing element
should read respectively $U_{a\pi} = (U_{a\pi})_{22}$ for 
$m_a < m_\pi$ and $U_{a\pi} = (U_{a\pi})_{21}$ for $m_a > m_\pi$. 

Despite all the mixing elements are continuous at the point $m_a=m_\pi$, see right panel in Fig.~\ref{fig:rotation}, 
at this point
there exists a 
discontinuity
proportional to $\epsilon \ll 1$ in the 
composition of the physical states, namely 
\begin{align}
    \pi_{\textrm{phys}} =\frac{\sqrt{1-\epsilon}}{\sqrt{2}}\left(a-\pi\right)\, , \quad a_{\textrm{phys}}=\frac{\sqrt{1+\epsilon}}{\sqrt{2}}\left(a+\pi\right)\,.
\end{align} 
For $m_a = m_\pi$, what we call ALP or pion is completely 
arbitrary, but  
as soon as we move away from this point the identification 
of the physical states in terms of ALP or pion becomes 
well defined. In practice, this is never a problem since 
the discontinuity will be small, as it is of order $\epsilon \lesssim 10^{-4}$, for realistic values of $f_a$. 

Away from the $m_a \approx m_\pi$ region, the transformation 
connecting the $a$ and $\pi$ to the physical fields, 
is usually approximated as 
\begin{align}
    \begin{pmatrix}
        a \\
        \pi
    \end{pmatrix} &= U_{a\pi} \begin{pmatrix}\pi_{\textrm{phys}} \\ a_{\textrm{phys}} \end{pmatrix} \approx \begin{pmatrix}
        \epsilon \frac{m_\pi^2}{m_a^2-m_\pi^2} & 1 \\
       1 & \epsilon \frac{m_a^2}{m_\pi^2-m_a^2}
    \end{pmatrix}\begin{pmatrix}\pi_{\textrm{phys}} \\ a_{\textrm{phys}} \end{pmatrix} \quad \textrm{for} \quad m_a \ll m_\pi \, , \\
    \begin{pmatrix}
        a \\
        \pi
    \end{pmatrix} &= U_{a\pi} \begin{pmatrix} a_{\textrm{phys}} \\ \pi_{\textrm{phys}} \end{pmatrix} \approx \begin{pmatrix}
      -1 &  \epsilon \frac{m_\pi^2}{m_a^2-m_\pi^2}  \\
        -\epsilon \frac{m_a^2}{m_\pi^2-m_a^2} & 1
    \end{pmatrix}\begin{pmatrix} a_{\textrm{phys}} \\ \pi_{\textrm{phys}} \end{pmatrix} \quad \textrm{for} \quad m_a \gg m_\pi \, , 
\end{align}
which are valid for $\abs{m_\pi^2-m_a^2}/\text{max}\{m_\pi^2,m_a^2\} \gg f_\pi/f_a$. 
Otherwise, the full formulae in terms of $U_{a\pi}$ should be employed. 

As a final remark, note that if we take the limit $f_\pi/f_a\to 0$ then Eq.~\eqref{eq:photon_effective} becomes
\begin{equation}
    \lim_{f_\pi/f_a\to 0} \frac{f_a}{f_\pi} (U_{a\pi})_{21} = \frac{c_{a\pi}}{2}\frac{m_a^2}{m_\pi^2-m_a^2}\, .
\end{equation}
This limit explains why the branching ratio into photons is independent of $\epsilon$. 
This is reflected in Fig.~\ref{fig:branchingratios}, where modifying the value of $f_a$ does not change the shape of the branching ratio to photons as long as $f_\pi\ll f_a$, making it independent of $\epsilon$.
 
\bibliographystyle{utphys}
\bibliography{ref}

\end{document}